\pdfoutput=1
\PassOptionsToPackage{hypertexnames=false, unicode, colorlinks=true, linkcolor=linkcolor, citecolor=linkcolor, filecolor=linkcolor, urlcolor=linkcolor, pdfusetitle}{hyperref}
\documentclass[10pt,aps,prd,amsmath,floats,floatfix, twocolumn,
superscriptaddress,nofootinbib,showpacs]{revtex4-1}

\usepackage[T1]{fontenc}
\usepackage[utf8]{inputenc}
\usepackage{lmodern}
\usepackage{orcidlink}
\usepackage{verbatim}
\usepackage[bottom]{footmisc}
\definecolor{linkcolor}{rgb}{0.0,0.3,0.5}
\usepackage[all]{hypcap}
\usepackage{graphicx}
\usepackage{xspace}
\usepackage{amssymb}
\usepackage[normalem]{ulem} %for \sout
\usepackage{bm} % boldmath
\usepackage{enumitem,amssymb}
\usepackage{lineno}
\usepackage{subfigure}
\usepackage{microtype}
\usepackage[english]{babel}
\usepackage{blindtext}
\usepackage{acro}  % Manager for acronyms
\DeclareAcronym{BSD}{
  short = BSD ,
  long  = Band Sampled Data
}

\DeclareAcronym{CR}{
  short = CR ,
  long  = Critical Ratio
}

\DeclareAcronym{FFT}{
  short = FFT ,
  long  = Fast Fourier Transform
}

\DeclareAcronym{CW}{
    short = CW ,
    long= continuous gravitational wave
}

\DeclareAcronym{GW}{
  short = GW ,
  long  = gravitational wave
}

\DeclareAcronym{BH}{
  short = BH ,
  long  = black hole
}

\DeclareAcronym{BSB}{
  short = BSB ,
  long  = Binary System Barycenter
}

\DeclareAcronym{SSB}{
  short = SSB ,
  long  = Solar-System Barycenter
}

\DeclareAcronym{O3}{
  short = O3 ,
  long  = the third observing run
}

\DeclareAcronym{O4a}{
  short = O4a ,
  long  = the first part of the fourth observing run
}

\DeclareAcronym{O4}{
  short = O4 ,
  long  = the fourth observing run
}

\DeclareAcronym{VBC}{
  short = VBC ,
  long  = ultralight vector boson cloud
}

\DeclareAcronym{LVK}{
  short = LVK ,
  long  = {LIGO Scientific Collaboration, Virgo Collaboration, and KAGRA Collaboration}
}

\DeclareAcronym{ASD}{
  short = ASD ,
  long  = {Amplitude Spectral Density}
}

\usepackage{lineno}

\newcommand{\scalarlimitsspin}{$[1.39, 6.94]\times 10^{-13}$}
\newcommand{\vectorlimitsspin}{$[0.32, 14.4]\times 10^{-13}$}
\newcommand{\vectorlimitssearch}{$[2.80, 3.95]\times 10^{-13}$}
\newcommand{\vectorlimitssearchone}{$[2.80, 3.26]\times 10^{-13}$}
\newcommand{\vectorlimitssearchtwo}{$[2.83, 3.95]\times 10^{-13}$}

\newif\ifshowcomments

\showcommentstrue
\newif\ifprintauthors
\printauthorstrue  
\begin{document}

%\preprint{}
%\linenumbers

\title{Constraints on ultralight bosons from merging binary and remnant black holes observed during the second and third parts of the fourth LIGO--Virgo--KAGRA observing run}

\ifprintauthors
  %% LVK authorlist in PRD format
%\documentclass[aps,prd,superscriptaddress]{revtex4-2}
%\begin{document}
%\title{LSC, Virgo and KAGRA February 2026 author list---LIGO-M2600028\\
%2026-04-15. PRD style}
\author{A.~G.~Abac\,\orcidlink{0000-0003-4786-2698}}
\affiliation{Max Planck Institute for Gravitational Physics (Albert Einstein Institute), D-14476 Potsdam, Germany}
\author{A.~Abe}
\affiliation{Department of Physics, Graduate School of Science, Osaka Metropolitan University, 3-3-138 Sugimoto-cho, Sumiyoshi-ku, Osaka City, Osaka 558-8585, Japan  }
\author{I.~Abouelfettouh}
\affiliation{LIGO Hanford Observatory, Richland, WA 99352, USA}
\author{F.~Acernese}
\affiliation{Dipartimento di Fisica ``E.R. Caianiello'', Universit\`a di Salerno, I-84084 Fisciano, Salerno, Italy}
\affiliation{INFN, Sezione di Napoli, I-80126 Napoli, Italy}
\author{K.~Ackley\,\orcidlink{0000-0002-8648-0767}}
\affiliation{University of Warwick, Coventry CV4 7AL, United Kingdom}
\author{A.~Adam}
\affiliation{OzGrav, University of Western Australia, Crawley, Western Australia 6009, Australia}
\author{S.~Adhicary\,\orcidlink{0009-0004-2101-5428}}
\affiliation{The Pennsylvania State University, University Park, PA 16802, USA}
\author{D.~Adhikari}
\affiliation{Max Planck Institute for Gravitational Physics (Albert Einstein Institute), D-30167 Hannover, Germany}
\affiliation{Leibniz Universit\"{a}t Hannover, D-30167 Hannover, Germany}
\author{R.~X.~Adhikari\,\orcidlink{0000-0002-5731-5076}}
\affiliation{LIGO Laboratory, California Institute of Technology, Pasadena, CA 91125, USA}
\author{V.~K.~Adkins}
\affiliation{Louisiana State University, Baton Rouge, LA 70803, USA}
% ...
\collaboration{The LIGO Scientific Collaboration, the Virgo Collaboration, and the KAGRA Collaboration}
%\begin{abstract}
%This is the LSC, Virgo and KAGRA February 2026 author list---LIGO-M2600028.
%\end{abstract}
%\maketitle
%\end{document}

\else
  \author{The LIGO Scientific
collaboration, the Virgo collaboration, the KAGRA collaboration}
\fi

%%%%%%%%%%%%%%%%%%%%%%%%%%%%%%%%%%%%%%%%%%%%%%%%%%%%%%%%%%%%%%%%%%%%%%%%%%%%%%%%
\begin{abstract}
We present constraints on ultralight bosons using binary black hole mergers observed in the second and third parts of the fourth LIGO--Virgo--KAGRA observing run.
Directed searches are conducted for long-transient gravitational waves from ultralight vector boson clouds around merger remnants, using a hidden-Markov-model (HMM) tracking scheme. We target the remnant black holes formed in the binary coalescences that produced GW250114 and GW250207.
We find no evidence for such signals from either target.
Estimating our search sensitivity at a threshold corresponding to a 1\% false alarm probability, we thus disfavor vector boson masses in the range of \vectorlimitssearch eV with greater than 90\% confidence.
In addition, we derive constraints on ultralight scalar and vector bosons from the inferred high spins of the constituent black holes in three binaries, using events GW240515, GW241113, and GW241225\_08. The excluded mass ranges in this approach depend on the assumed black-hole ages. At $10^5$ years, corresponding to typical dynamically formed binaries, we exclude scalar and vector bosons in the ranges \scalarlimitsspin eV and \vectorlimitsspin eV at 90\% confidence, respectively.

%\textit{Draft revision: \IfFileExists{./gitID.txt}{git hash \input{gitID.txt}}{Compiled on Overleaf} (\today)}
\end{abstract}

%\keywords{keyword1,keyword2}

\maketitle
%%%%%%%%%%%%%%%%%%%%%%%%%%%%%%%%%%%%%%%%%%%%%%%%%%%%%%%%%%%%%%%%%%%%%%%%%%%%%%%%
\section{Introduction}
\label{Sec:intro}
%%%%%%%%%%%%%%%%%%%%%%%%%%%%%%%%%%%%%%%%%%%%%%%%%%%%%%%%%%%%%

Ultralight bosons arise in a variety of extensions of the Standard Model and 
can provide viable dark matter candidates~\cite{Bertone2018_1, Bertone2018_2, Oks2021, Baryakhtar2025}.
Motivated examples include axion-like particles and other spin-0 fields~\cite{Arvanitaki2010, Freitas2021}, as well as hidden-sector vector (spin-1) bosons such as dark photons~\cite{Holdom1986, Goodsell2009, Jaeckel2010, Essig2013, Agrawal2020, Fabbrichesi2021}.
%and tensor (spin-2)~\cite{Clifton2012, Babichev2016, Babichev2016_2, Aoki2016, Aoki2018, Yusuke2023} bosons.
Even if these particles interact only through gravity, rotating black holes can trigger superradiant instabilities that populate bound states of the bosonic field. The resulting boson clouds can grow to macroscopic occupation numbers and act as long-lived sources of nearly monochromatic gravitational radiation, potentially observable with ground-based detectors~\cite{Arvanitaki2011, Yoshino2014, Yoshino2015, Arvanitaki2015, Arvanitaki2017, Brito2017, Brito2017_2, Baryakhtar2017, Chan2022, Cardoso2018, Baumann2019, Hannuksela2019, Zhang2019, East2017, East2017_2, East2018, Siemonsen2020, Brito2020}.

A rotating black hole becomes superradiantly unstable to an ultralight bosonic field of mass $m_b$ when the condition
\begin{equation}    
     \omega_b < m \Omega_H
    \label{eqn:superrad_condition}
\end{equation}
is satisfied. Here, $\omega_b \approx m_b c^2 / \hbar$ is the angular frequency of the bosonic field, $m$ is the azimuthal quantum number of the bosonic bound state, and $\Omega_H$ is the angular frequency of the black hole horizon. 
The instability is most efficient when the boson Compton wavelength $\lambda$ is comparable to the black hole gravitational radius $r_g$, causing the occupation number of the bound state to grow exponentially.
The instability extracts energy and angular momentum from the black hole into the bosonic field, leading to a coupled evolution in which the black hole spins down while the cloud grows. If gravitational interactions dominate, this process results in the amplification of the field into a macroscopic boson cloud, which can extract up to $\sim 10\%$ of the initial black hole mass~\cite{Zel'Dovich1971, Starobinskii1973, Detweiler1980, Bekenstein1973, Dolan2007, Arvanitaki2011, East2017_2, East2018, Herdeiro2022}.

This extraction continues until the superradiance condition in Eq.~\eqref{eqn:superrad_condition} begins to saturate ($\omega_b \approx m \Omega_H$), at which point the black hole has spun down sufficiently to shut off the instability.
When it satisfies the superradiant condition, the $m=1$ mode is typically the fastest growing and most relevant for a newly formed black hole. Even after the saturation of this mode, higher modes can continue to grow through superradiance and further spin down the black hole.
The boson cloud then depletes through gravitational-wave emission at approximately twice the oscillation frequency, with a small positive frequency drift as the binding energy decreases~\cite{East2017_2, East2018, Baryakhtar2017, Isi2019, Siemonsen2020, Baryakhtar2021, Siemonsen2023_2, May2024}. 
For stellar- and intermediate-mass black holes and boson masses in the range $\sim 10^{-14}$--$10^{-11}~\mathrm{eV}$~\cite{Arvanitaki2015, Brito2017_2}, this emission falls within the sensitive band of ground-based gravitational-wave detectors. Current and future detector networks~\cite{aLIGO, aVirgo, KAGRA,GW_observation_prospects,ET:2025xjr, Evans2021} therefore provide a promising avenue to probe ultralight bosons.
In addition, measurements of black hole spins offer a complementary probe, as superradiance-induced spin-down can imprint observable signatures on the spin distribution of merging systems~\cite{Arvanitaki2011, Arvanitaki2015, Brito2017_2, East2018, Cardoso2018}.

Constraints on ultralight scalar~\cite{Arvanitaki2015, Ng2021, Cardoso2018, Brito2017_2, aswathi2025, caputo2025, Abac_2025_GW241011_241110,Ning:2026ebu,Witte2025} and vector~\cite{Baryakhtar2017, Cardoso2018, aswathi2025, Abac_2025_GW241011_241110} bosons have been derived using black hole spin measurements, although these are limited by uncertainties in inferred spins and masses arising from modelling assumptions and observational systematics~\cite{Mummery:2025tvq, Reynolds:2019uxi, McClintock:2013vwa}.
%\blue{Recent work has refined spin-based axion constraints by incorporating self-interactions and the coupled evolution of higher bound states within a Bayesian framework, while assessing uncertainties in X-ray spin measurements~\cite{Witte2025}.}
Searches for continuous gravitational waves, including all-sky and Galactic-center targeted analyses, also constrain scalar bosons under assumptions about the black hole population (mass, spin, and age distributions), disfavoring masses of order $\sim 10^{-13}$--$10^{-12}$~eV~\cite{O3_all-sky_scalar_bosons, Palomba2019, Dergachev2019, O3_galactic_center, Zhu2020,Ellis2026}. Similarly, stochastic background searches for populations of black holes with scalar~\cite{Tsukada2019, Yuan2022, Ellis2026} and vector~\cite{Tsukada2021} clouds disfavor boson masses around $\sim 10^{-13}$~eV, subject to population modelling assumptions.
%\blue{Population studies have also forecast the sensitivity of current and future detectors to Galactic continuous-wave signals and the extragalactic stochastic background from axion clouds, accounting for uncertainties in black hole populations~\cite{Ellis2026}.}
Directed searches targeting Cygnus~X-1 exclude scalar and vector bosons in the range $\sim [0.6, 1]\times 10^{-12}$~eV and $\sim [0.9,\,1.6]\times 10^{-13}$~eV, respectively, depending on the assumed black hole age and initial spin~\cite{O2_CygnusX1_scalar_bosons, Collaviti2024, LIGOScientific:2025csr}.  
A continuous-wave search targeting pulsar-catalog sources as candidate luminous dark-photon clouds constrained kinetically mixed dark photons under assumptions about the black hole population and electromagnetic emission~\cite{Mirasola2025}.
While these results overlap with the parameter space considered in this work, they rely on astrophysical assumptions regarding black hole populations, evolution, and ages; as such, their implications for ultralight bosons in these ranges should be interpreted with caution.  
In addition to indirect searches, ground-based interferometers have been used for direct searches for ultralight scalars~\cite{Vermeulen2021, Gottel2024, Aiello2022, LIGOScientific:2025ttj} and vectors~\cite{direct_dark_matter_LIGO_Virgo, direct_dark_matter_KAGRA, LIGOScientific:2025ttj}. Future observations may also probe the impact of boson clouds on binary black hole inspirals for certain masses and field strengths~\cite{Yang2018, Baumann2019, Choudhary2021,Roy:2025qaa,Tomaselli2024}.
%\blue{Cloud-binary resonances can modify orbital eccentricities and orientations, potentially leaving observable imprints on binary populations even after the clouds have been depleted~\cite{Tomaselli2024}.}

In this study, we present results from two complementary approaches to probe ultralight bosons with gravitational-wave observations. First, we perform a directed search for long-duration transient gravitational-wave signals (on a timescale of hours to months) from vector boson clouds around merger-remnant black holes using a Hidden Markov Model (HMM)~\cite{Jones2023,Jones2025}. 
Targeting newly formed black holes with well-characterized properties enables robust constraints with reduced reliance on assumptions about the underlying black hole population and age.
Although scalar boson clouds around merger remnant black holes can also generate long-duration gravitational waves~\cite{Isi2019,Ghosh:2018gaw}, their strain amplitudes are generally expected to lie below the sensitivity of current-generation detectors. 
Vector boson clouds, by contrast, are expected to produce stronger and shorter-lived emission, making them more favorable targets for directed searches with current-generation detectors~\cite{Jones2023}.
This approach follows Refs.~\cite{Jones2023,Jones2025} and was previously applied to O4a events~\cite{LIGOScientific:2025csr}, constituting the first directed gravitational-wave search for ultralight vector bosons targeting merger remnants. 
In O4a, the merger remnants of GW230814\_230901 and GW231123\_135430 (hereafter GW230814
and GW231123) were targeted, yielding weakly disfavored vector boson masses of $[0.94,\,1.08]\times 10^{-13}~\mathrm{eV}$ and $[2.75,\,3.28]\times 10^{-13}~\mathrm{eV}$ at $\sim 30\%$ confidence. Reference~\cite{LIGOScientific:2025csr} also performed a continuous-wave search using the Band Sampled Data framework, targeting an older galactic black hole in Cygnus~X-1 and excluding vector boson masses in the range $[0.85,\,1.59]\times 10^{-13}~\mathrm{eV}$ at 95\% confidence, assuming an initial black hole spin $\gtrsim 0.5$.
In this work, we extend the HMM analysis to two especially well-suited events from the second and third parts of the fourth LVK observing run (O4b~\cite{GWTC-5} and O4c~\cite{GW250207}), focusing on merger remnants formed in binary black hole coalescences. For the remnant search targeting the early post-merger epoch, we consider only the fastest-growing mode with azimuthal number $m=1$.
This study demonstrates that directed searches targeting these events can already strongly disfavor vector boson masses within a relatively narrow mass range set by the remnant properties.

Second, we derive constraints from pre-merger black hole spin measurements following Ref.~\cite{aswathi2025}. This approach identifies regions of parameter space incompatible with efficient superradiant spin-down, thereby constraining the allowed boson masses. 
In this spin-based analysis, we consider all relevant azimuthal modes that can grow within the assumed black-hole lifetime.
Previous studies have shown that events such as GW190517\_055101, GW231123, and GW241110\_124123 disfavor scalar and vector bosons over the mass ranges $[0.55,\,26]\times 10^{-13}~\mathrm{eV}$ and $[0.11,\,53]\times 10^{-13}~\mathrm{eV}$, respectively, at a 90\% confidence level,  assuming black hole lifetimes $\gtrsim 10^5~\mathrm{yr}$~\cite{aswathi2025,Abac_2025_GW241011_241110}. 
We extend this analysis to high-spin events observed during O4b and reported in GWTC-5~\cite{GWTC-5-intro,GWTC-5-method,GWTC-5}, providing a complementary and relatively model-independent probe of ultralight bosons. 

The paper is organized as follows. In Sec.~\ref{Sec:sources}, we identify the target sources used in both analyses. In Sec.~\ref{Sec: HMM_search}, we present the HMM-based search for long-duration transient gravitational-wave signals from ultralight vector boson clouds, including the search methodology, configuration, candidate follow-up procedure, and the resulting disfavored parameter space. In Sec.~\ref{Sec:spin_measurements_constraints}, we describe the complementary analysis based on black hole spin measurements, outlining the method and presenting the corresponding results. Finally, we summarize our findings in Sec.~\ref{Sec:conclusions}.

%%%%%%%%%%%%%%%%%%%%%%%%%%%%%%%%%%%%%%%%%%%%%%%%%%%%%%%%%%%%%%%%
\section{Target Sources \label{Sec:sources}}
%%%%%%%%%%%%%%%%%%%%%%%%%%%%%%%%%%%%%%%%%%%%%%%%%%%%%%%%%%%%%%%
In this section, we describe the black hole systems targeted by the two complementary analyses presented in this work and summarize the gravitational-wave and superradiance parameters relevant to each.
%For merger remnants, the expected signal properties— including the cloud growth timescale, gravitational-wave emission frequency, frequency evolution, and characteristic strain—are computed using the SuperRad waveform model. And for highly spinning pre-merger systems, we use the same framework to determine the maximum black hole spin permitted in the presence of superradiant instabilities, thereby identifying regions of boson mass–coupling parameter space that are excluded by the observed component spins.

\subsection{Merger remnant black holes \label{Subsec: HMM_sources }}

The merger remnants analyzed in this work are selected based on their expected detectability under the ultralight vector boson cloud scenario. 
For a subset of compact binary coalescence candidates with a high signal-to-noise ratio (SNR) identified in O4b and O4c, we use the median estimated source properties from GWTC-5 for O4b and from the preliminary parameter-estimation (PE) results available at the time for O4c~\cite{GWTC-5, Moe:2014GraceDB,GWTC-5-method,GWTC5_PE_Zenodo_V9_Part1,GWTC5_PE_Zenodo_V9_Part2,GW250114,GW250114-comp,GW250207}.
For each event, we estimate the maximum detectable luminosity distance, referred to as the horizon distance, for boson-cloud gravitational-wave emission from a black hole with the corresponding median remnant mass and spin, and compare it with the median estimated luminosity distance $D_{\mathrm{L}}$.\footnote{Here, the horizon distance refers to the reach for gravitational-wave emission from the vector-boson cloud around the remnant black hole, rather than for the binary merger itself. The evaluation using median values is intended only as a representative criterion for target selection. The selection is deliberately permissive rather than strict, since the horizon distance is, by definition, an optimistic estimate of the detectable reach.}
The horizon distances are estimated empirically through simulations over the remnant mass and spin parameter space, as described in Ref.~\cite{Jones2023}.
Systems with $D_{\mathrm{L}}$ smaller than the horizon distance are expected to lie within the sensitivity reach of the HMM search. 
Two merger remnants satisfy this criterion with high confidence in the preliminary selection: GW250114\_082203 in O4b and GW250207\_115645 in O4c (hereafter GW250114 and GW250207), among the loudest gravitational-wave signals observed to date~\cite{GWTC-5}.
Using the final PE results~\cite{GW250114,GW250114-comp,GW250207}, we verify that the luminosity distances of both systems lie well within their corresponding horizon distances. The estimated horizon distances are approximately 574~Mpc for GW250114 and 592~Mpc for GW250207, substantially larger than their median luminosity distances of 403~Mpc and 188~Mpc, respectively.

\begin{figure*}[t]
 \centering
 \subfigure[]{\includegraphics[width=0.45\linewidth]{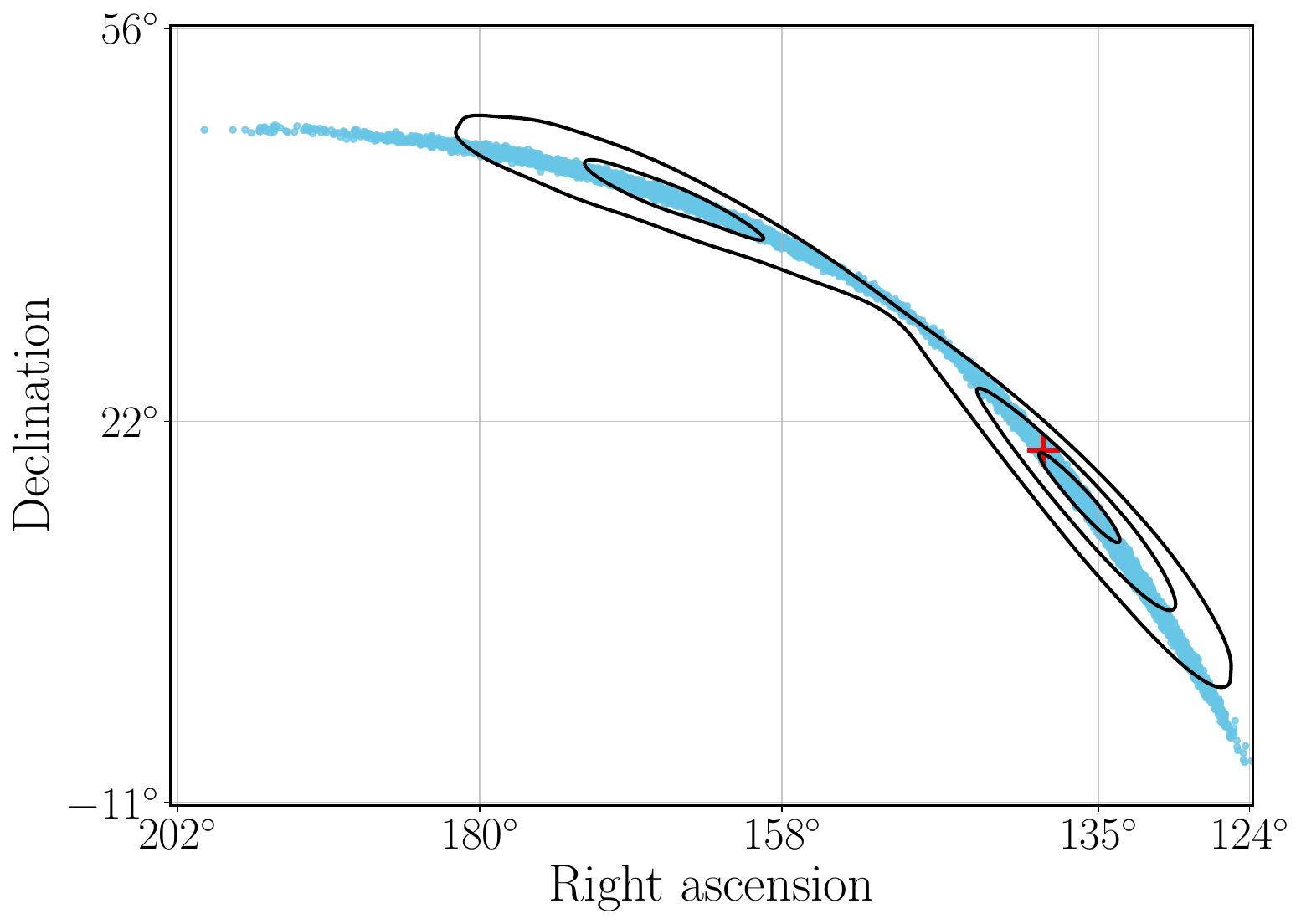} \label{fig:S250114_skylocalisation}}\qquad
  \subfigure[]{\includegraphics[width=0.45\linewidth]{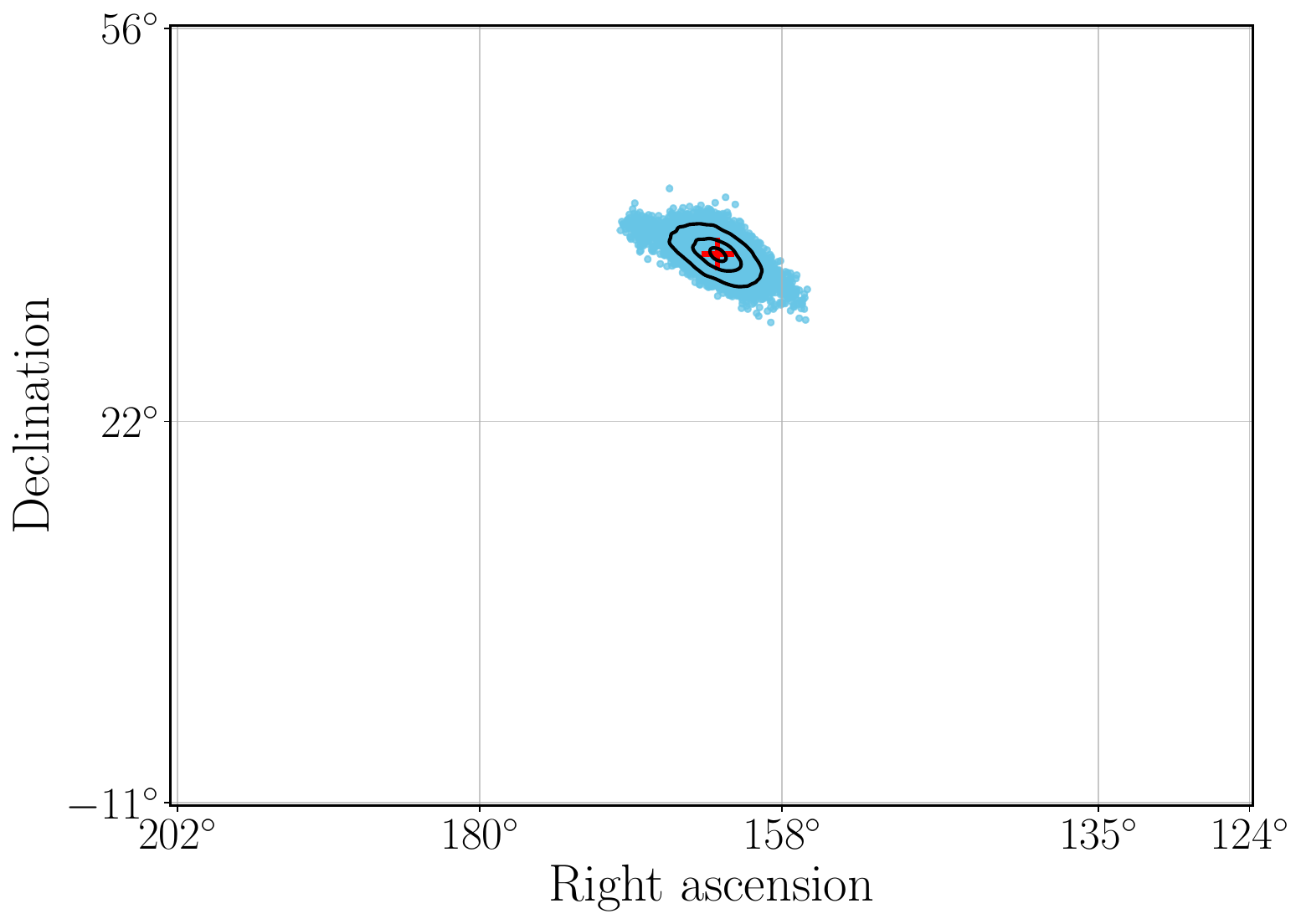}\label{fig:S250207_skylocalisationq}} 
  \caption{Joint right ascension and declination posterior distribution for GW250114 (left) and GW250207 (right). The blue points represent sky-position samples from the full posterior distribution. The black contours show the 90\%, 50\%, and 10\% credible intervals. The plus markers indicate the sky positions targeted in the search for each remnant black hole (median values of the distributions).
  }
  \label{fig:skylocalisation}
\end{figure*}

The median parameters for each remnant and the 90\% credible sky area are summarized in Table~\ref{tab:hmm_targets}, where $M$ and $\chi$ denote the mass and dimensionless spin of the black hole prior to superradiance, $D_{\mathrm{L}}$ is the luminosity distance, and $\iota$ is the inclination angle. The sky localizations for both events are shown in Fig.~\ref{fig:skylocalisation}.
Although the two events have different localization uncertainties, both localizations are sufficiently compact for the search considered here (see relevant search configuration in Sec.~\ref{Subsec:search_configuration}).
The PE results used for GW250114 and GW250207 are obtained with the NRSur7dq4 waveform model~\cite{Varma2019}. This model is employed because it interpolates directly between numerical relativity simulations within its training domain, thereby reducing additional waveform-modeling assumptions, and is among the most reliable waveform models for relatively heavy binary black hole systems and their remnant properties. 

%For the remaining events considered in O4b, we employ the latest available PE results at the time of this analysis. These PE are preliminary and will be updated once the final parameter estimation analyses are completed.

\begin{table*}[t]
\centering
\caption{Properties of the merger remnants targeted in the HMM search.}
\label{tab:hmm_targets}
\renewcommand{\arraystretch}{1.5} 
\setlength{\tabcolsep}{8pt}
\begin{tabular}{lccccc}
\hline\hline
Source  $^{\rm a, b}$ & $M_i$ [M$_\odot$]  & $\chi_i$ &  $D_{\mathrm{L}}$ [Mpc] & $\cos \iota$ & 90 \% credible sky area [deg$^2$] \\ %RA [rad] & Dec [rad] \\
\hline
GW250114 & $62.7^{+1.0}_{-1.1}$ & $0.68^{+0.01}_{-0.01}$ & $403^{+74}_{-70}$ & $0.64^{+0.18}_{-1.4}$ & 254.01\\% &$2.42^{+0.65}_{-0.14}$ &$0.34^{+0.44}_{-0.24}$ \\
GW250207 & $62.8^{+0.9}_{-1.4}$ & $0.69^{+0.01}_{-0.01}$ & $188^{+118}_{-52}$ &$0.38^{+0.30}_{-0.1}$ & 18.46 \\
 % &$2.84^{+0.04}_{-0.04}$ &$0.63^{+0.03}_{-0.03}$ \\
\hline\hline
\end{tabular}
\footnotetext[1]{We use $M_i$ and $\chi_i$ to denote the mass and dimensionless spin of the black hole before superradiance occurs, corresponding to the final black hole mass $M_f$ and spin $\chi_f$ inferred from the compact-binary-coalescence PE. We avoid using $M_f$ and $\chi_f$ here to prevent confusion with the black hole mass and spin after superradiance.}
\footnotetext[2]{The two remnant black holes have similar masses and spins; however, differences in detector duty cycles and data quality lead to complementary results from these two targets (see Sec.~\ref{Sec: HMM_search}).}
\end{table*}

\subsection{Pre-merger black holes with high spin\label{Subsec: spin_constrain_sources }}

In addition to merger remnants, we target compact binary coalescences exhibiting highly spinning constituent black holes prior to merger. The presence of large dimensionless spins provides a complementary probe of ultralight bosons through black hole superradiance: if a boson with a given mass exists, rapidly rotating black holes are expected to spin down over astrophysical timescales. Therefore, the observation of high spins can be used to exclude regions of the boson mass parameter space.

Events considered in this analysis are selected based on (i) evidence for high constituent spins and (ii) well-constrained spin posteriors. Specifically, we require that the median effective inspiral spin satisfies $|\chi_{\rm eff}| > 0.2$, and that at least one component black hole has a median dimensionless spin magnitude greater than 0.5, with its 5th percentile exceeding 0.3. The additional requirement on the 5th percentile ensures that the inferred high spin is not driven solely by the posterior median, and disfavors systems whose posterior distributions retain substantial support for low spins. 
Applying these criteria to the O4b catalog (reported in GWTC-5~\cite{GWTC-5}) yields three events of particular interest: GW240515\_005301, GW241113\_163507, and GW241225\_082815 (hereafter GW240515, GW241113, and GW241225\_08\footnote{We define the longer abbreviated name GW241225\_08 for GW241225\_082815 because GWTC-5 contains two events from 2024 December 25.}). 

We use PE results obtained with the \texttt{bilby} framework~\cite{Ashton:2018jfp,Romero-Shaw:2020owr} and the \texttt{dynesty} sampler~\cite{Speagle:2019ivv}, employing the NRSur7dq4 waveform model~\cite{Varma2019} for GW240515 and GW241225\_08 and the IMRPhenomXPHM-SpinTaylor waveform model~\cite{Colleoni2025,Pratten:2020ceb} for GW241113.\footnote{NRSur7dq4 is used where applicable because it provides reliable parameter estimates for systems in the higher-mass range~\cite{Varma2019}.
For the lower-mass event GW241113, which lies outside the NRSur7dq4 coverage, we instead use IMRPhenomXPHM-SpinTaylor~\cite{Colleoni2025,Pratten:2020ceb}.}
NRSur7dq4 was trained on numerical-relativity simulations with component spin magnitudes $\chi_{1,2}\leq0.8$~\cite{Varma2019}.
It has also been used in parameter-estimation analyses allowing larger spins, which require extrapolation beyond the training domain~\cite{GWTC-4_results,GWTC-5}.
For events with posterior support above this threshold, particularly GW241225\_08, this extrapolation represents a potential source of waveform-model uncertainty in the inferred spins and resulting superradiance constraints.
From these PE results, we extract the primary and secondary masses ($m_1$, $m_2$), the corresponding dimensionless spin magnitudes ($\chi_1$, $\chi_2$), the luminosity distance $D_{\mathrm{L}}$, and the network SNR. The median values of these parameters are summarized in Table~\ref{tab:highspin_targets}.

\begin{table*}[t]
\centering
\caption{Median source parameters of the high-spin pre-merger systems selected from O4b.}
\label{tab:highspin_targets}
\renewcommand{\arraystretch}{1.5} 
\setlength{\tabcolsep}{8pt}
\begin{tabular}{lcccccc}
\hline\hline
Source & $m_1$ [M$_\odot$] & $m_2$ [M$_\odot$] & $\chi_1$ & $\chi_2$ & $D_{\mathrm{L}}$ [Mpc] & SNR \\
\hline
GW240515 &$36.8^{+11.6}_{-10.1}$  & $18.5^{+8.7}_{-4.5}$ & $0.71^{+0.25}_{-0.38}$ &  $0.56^{+0.39}_{-0.50}$ & $3794^{+1840.21}_{-1876.50}$ & $10^{+1.7}_{-1.7}$  \\
GW241113 & $19.3^{+5.4}_{-3}$ & $14.3^{+2.7}_{-3.3}$ &$0.80^{+0.17}_{-0.38} $ & $0.56^{+0.39}_{-0.49}$ & $1355.44^{+666.50}_{-631.87}$ & $ 11.7^{+1.7}_{-1.7}$ \\
GW241225\_08 & $58.5^{+11.1}_{-9.9}$ & $36.3^{+9.7}_{-11.1}$ &$0.84^{+0.14}_{-0.41} $ & $0.48^{+0.46}_{-0.43}$ & $1972^{+1166.74}_{-954.62}$ & $ 17.7^{+1.7}_{-1.6}$ \\
\hline\hline
\end{tabular}
\end{table*}

%%%%%%%%%%%%%%%%%%%%%%%%%%%%%%%%%%%%%%%%%%%%%%%%%%%%%%%%%%%%%%%%%%%%%%%

\section{Searches targeting merger remnants \label{Sec: HMM_search}}

\subsection{Method \label{Subsec:search_method}}

The directed searches are performed using the HMM tracking pipeline described in Refs.~\cite{Jones2023,Jones2025} and applied previously in O4a~\cite{LIGOScientific:2025csr}. 
Although the signal evolution is well modeled for fixed source parameters, the parameter space in boson mass and remnant black-hole properties is too large for a fully coherent matched-filter search over signal durations ranging from hours to months. The HMM approach therefore provides a computationally tractable way to track the expected frequency evolution while maintaining good coverage of the relevant parameter space.
The pipeline computes a coherent multi-detector $ \mathcal{F}$-statistic~\cite{JKS,Cutler2005,Prix2010Fstat} by coherently combining short Fourier transforms over segments of
duration $T_{\rm coh}$, referred to as the coherent integration time.
The HMM then tracks the signal frequency evolution across segments  via the Viterbi algorithm~\cite{Viterbi1967}, without requiring
phase coherence between segments. 
Longer $T_{\rm coh}$ improves sensitivity when the signal remains well
matched to the coherent template, but also narrows the frequency bins.
Rapid frequency evolution therefore limits the coherent integration
time, as excessive drift within a segment spreads the signal power
across multiple bins and reduces sensitivity.
We choose $T_{\rm coh}$ (and corresponding frequency resolution) based on the maximum expected frequency derivative of the signal, ensuring that the signal remains approximately monochromatic within each segment and evolves by at most one frequency bin between adjacent segments. 
The significance of candidate tracks is quantified using a log-likelihood-based detection statistic $\mathcal{L}$, normalized by the number of segments $N_T$ as $\bar{\mathcal{L}} \equiv \mathcal{L}/N_T$, allowing for consistent comparison across different search configurations. 
We refer the reader to Refs.~\cite{Jones2023,Jones2025,LIGOScientific:2025csr} for a detailed description of the pipeline and its implementation.
For the synthetic-signal simulations in this study, we compute vector-boson signal waveforms using the superradiance waveform model \texttt{SuperRad}~\cite{May2024,Siemonsen2023}, and generate the simulated data with the \texttt{simulateCW} module of the \texttt{LALPulsar} package in \texttt{LALSuite}~\cite{lalsuite,swiglal}.

As the remnant black holes are relatively young and isolated, the corresponding signals are expected to last hours to months and can exhibit rapid frequency evolution. We therefore employ the HMM search method, which is well-suited to signals with significant frequency drift and associated uncertainties.

%%%%%%%%%%%%%%%%%%%%%%%%%%%%%%%%%%%%%%%%%%%%%%%%%%%%%%%%%%%%%%%%%%

\subsection{Signal parameter space\label{Subsec:signal_paras}} 

We specialize to vector bosons and denote their mass by $m_V$.
%Although scalar boson clouds around merger remnant black holes can also generate long-duration gravitational waves~\cite{Isi2019}, their strain amplitudes are generally expected to lie below the sensitivity of current-generation detectors. 
%Vector boson clouds, by contrast, are expected to produce stronger and shorter-lived emission, making them more favorable targets for directed searches with current-generation detectors~\cite{Jones2023}.
Using the \texttt{SuperRad} waveform model~\cite{May2024,Siemonsen2023} together with the median black hole mass and spin estimates listed in Table~\ref{tab:hmm_targets}, we determine the optimal boson mass $m_V^{\rm opt}$ that maximizes the superradiant instability for the corresponding black hole. This $m_V^{\rm opt}$ corresponds to the vector-boson mass for which the gravitational-wave strain amplitude is maximized at the reference time $t_{\rm ref}$, when the $m=1$ mode reaches saturation.
For GW250114 and GW250207, we evaluate the representative boson masses $m_V^{\mathrm{opt}} = 3.5 \times 10^{-13}~\mathrm{eV}$ and $3.6 \times 10^{-13}~\mathrm{eV}$, respectively, corresponding to the median remnant black hole parameters (see Table~\ref{tab:signal_parameters}).
Using these values $m_V^{\mathrm{opt}}$ and the median remnant black hole parameters, we compute the corresponding gravitational-wave signal properties during the early post-merger epoch with the \texttt{SuperRad} model, assuming the dominant $m=1$ superradiant mode. The resulting signal parameters, listed in Table~\ref{tab:signal_parameters}, are defined as follows. The detector-frame growth and depletion timescales, $\tau_{\rm growth}^{\rm det}$ and $\tau_{\rm GW}^{\rm det}$, describe the evolution of the boson cloud. The former is the inverse growth rate, while the latter corresponds to the gravitational-wave emission timescale. The quantities $h_0^{\rm eff}$, $f_0$, and $\dot{f}_0$ denote the effective strain, detector-frame signal frequency, and frequency drift at $t_{\rm ref}$. The effective strain amplitude seen by the detector is given by $h_0^{\rm eff} = h_0\,2^{-1/2} \{[(1+\cos^2{\iota})/2]^2 + \cos^2{\iota}\}^{1/2}$~\cite{Sun2018}, where $h_0$ is the gravitational-wave strain on Earth from a source at luminosity distance $D_{\mathrm{L}}$, and $\iota$ is the inclination angle.\footnote{Here, we use $\iota$, the angle between the binary orbital angular momentum and the line of sight, instead of the viewing angle $\theta_{JN}$, defined as the angle between the total angular momentum of the binary and the line of sight. For GW250114 and GW250207, the posterior distributions of these two angles are similar because both systems have nearly equal component masses, small component spins, and no evidence for significant spin precession.}
The detailed signal evolution model is described in Appendix~\ref{app:vector_waveform}.
The signal properties evaluated at $m_V^{\rm opt}$ with the median remnant parameters in Table~\ref{tab:signal_parameters} provide representative examples of the expected signals for the two targets. The search itself is not restricted to these values: it covers a range of vector boson masses and accounts for uncertainties in the remnant black hole properties through their posterior distributions (see Sec.~\ref{Subsec:search_configuration}).

We consider a search mass range of $[0.6,\,1.1]\,m_V^{\mathrm{opt}}$, over which the method is expected to be sensitive, as determined empirically in Ref.~\cite{Jones2025}.
This corresponds to $[2.13,\,3.90] \times 10^{-13}~\mathrm{eV}$ for GW250114 and $[2.18,\,3.99] \times 10^{-13}~\mathrm{eV}$ for GW250207. 
The range includes vector masses that, while having lower strain amplitude,  can yield longer-lived gravitational-wave emission, allowing longer coherent integration times and, in some cases, greater sensitivity than at $m_V^{\mathrm{opt}}$.

\begin{table*}[t]
\centering
\caption{Estimated gravitational-wave signal parameters (at the reference time) for the merger remnant targets using the median black hole parameters and their respective $m_V^\mathrm{opt}$ values.}
\label{tab:signal_parameters}
\renewcommand{\arraystretch}{1.5} 
\setlength{\tabcolsep}{8pt}
\begin{tabular}{lcccccc}
\hline\hline
Source & $m_V^\mathrm{opt}$ [$10^{-13}$ eV] & $\tau_\mathrm{growth}$ [h] & $\tau_\mathrm{GW}$ [h] & $h_0^\mathrm{eff}$ [$10^{-24}$] & $f_0$ [Hz] & $\dot{f}_0$ [Hz s$^{-1}$] \\
\hline
GW250114 &3.5 &8.8  &32.6 &0.381 &156.2 & 4.83 $ \times 10^{-7}$ \\
GW250207 &3.6 &7.2 & 24.8 &0.606 &166.1 &7.32 $ \times 10^{-7}$ \\
\hline\hline
\end{tabular}
\end{table*}

%%%%%%%%%%%%%%%%%%%%%%%%%%%%%%%%%%%%%%%%%%%%%%%%%%%%%%%%%%%%%%%%%%
\subsection{Data set and search configuration\label{Subsec:search_configuration}} 

In this section, we describe the data set and analysis setup used to perform a directed search for long-transient gravitational-wave emission from the merger remnants of GW250114 and GW250207.

We analyze calibrated, analysis-ready strain data from the Advanced LIGO detectors~\cite{aLIGO} at Hanford (H1) and Livingston (L1) during the O4b and O4c observing period~\cite{Capote2025, Soni2025, Ganapathy2023, Jia2024,segments_O4}. The strain is produced by the low-latency calibration pipeline (C00), with gating applied to suppress transient noise artifacts and improve the robustness of the HMM search~\cite{o4gating}. We use the gated strain channels \texttt{H1:GDS-CALIB\_STRAIN\_CLEAN\_GATED\_G02} and \texttt{L1:GDS-CALIB\_STRAIN\_CLEAN\_GATED\_G02}~\cite{O4cal, Wade2025, Sun2020, Viets2018, Viets_thesis, Soni2025, Vajente2020, open_data_O4b, o4gating}
%accessed via the frame files listed in H1\_HOFT\_C00\_GATED\_G02-O4-full, L1\_HOFT\_C00\_GATED\_G02-O4-full. 

Within the vector-boson mass search ranges of $[2.13,\,3.90]\times 10^{-13}~\mathrm{eV}$ for GW250114 and $[2.18,\,3.99]\times 10^{-13}~\mathrm{eV}$ for GW250207, each boson mass maps to a distinct gravitational-wave emission frequency, corresponding to detector-frame frequency ranges of $[92,\,175]~\mathrm{Hz}$ and $[96,\,185]~\mathrm{Hz}$, respectively.
We divide these ranges into $1~\mathrm{Hz}$ sub-bands over which the search is performed.
Given the relatively precise sky localization of these sources (see Table~\ref{tab:hmm_targets}), we adopt the median right ascension and declination shown in Fig.~\ref{fig:skylocalisation} as fixed sky positions for the search. 
Although nearly monochromatic, the targeted signals are long transients lasting hours to months, rather than persistent continuous-wave emission.
Previous injection studies have shown that, for well-localized sources, using a single representative sky position is sufficient for this type of search because the targeted signals are relatively short and the sensitivity is only weakly affected by offsets within the localization region~\cite{Jones2023,Jones2025,LIGOScientific:2025csr}.
While the localization uncertainty differs between the two events, both are sufficiently well constrained to justify using a fixed sky position, without incurring the substantial computational cost of searching over a sky grid. Any residual sensitivity loss due to offsets from the true source position is consistently reflected in the sensitivity estimates of Sec.~\ref{Subsec:directed_searches_constraints}, where signal recovery is evaluated using the same fixed sky-position setup as in the search.

Following the guidelines in Ref.~\cite{Jones2023}, we randomly draw 200 samples from the multidimensional posterior distribution of the remnant black hole. 
For each sample of black hole parameters and each boson mass considered in the search, we determine the optimal search configuration $(t_{\rm start}, T_{\rm coh}, T_{\rm obs})$, where $t_{\rm start}$ is the GPS start time of the search, corresponding to the time at which the cloud reaches saturation, and $T_{\rm obs}$ is the full search duration. We set $T_{\rm obs}=\tau_{\rm GW}^{\rm det}$, but typically cap it at 180 days to limit the computational cost of searching for sub-optimally matched boson masses~\cite{Jones2021,LIGOScientific:2025csr}. 
This cap, however, does not limit any of the search configurations used in this analysis, as detailed below.
We then construct 11 representative search configurations by independently taking the 2nd, 10th, 20th, 30th, 40th, 50th, 60th, 70th, 80th, 90th, and 98th percentiles of the resulting marginal distributions of $t_{\rm start}$, $T_{\rm coh}$, and $T_{\rm obs}$.
These configurations, summarized in Table~\ref{tab:config} for each event, are used in the HMM search, with the detector noise amplitude spectral densities estimated from the data over the corresponding coherent segments and searched frequency bands as part of the $\mathcal{F}$-statistic calculation~\cite{Prix2010Fstat}.
No single configuration is optimal over the full parameter space because of the uncertainties in the remnant black-hole properties and the unknown boson mass, while using an individually optimized configuration for each parameter set would be computationally infeasible. We therefore use multiple configurations to balance broad coverage with search depth, improving sensitivity across the physically plausible signal parameter space while remaining computationally tractable.
For most systems drawn from the remnant posterior distribution, the signal is expected to be recoverable by more than one configuration, making this percentile spacing a conservative choice.

We exclude the configuration corresponding to the 98th percentile (configuration 11) from the analysis.
This choice is motivated by the interplay between the expected signal persistence timescales and the O4c observing schedule. In particular, the longest-lived signals may extend to 13 May 2025 for GW250114 (GPS time $1431151905~\mathrm{s}$) and 11 May 2025 for GW250207 (GPS time $1431017883~\mathrm{s}$). These intervals overlap with a scheduled commissioning break in O4c from 1 April 2025 to 11 June 2025 (GPS: 1427554818--1433689218), during which no observational data are available for coherent tracking over the full signal duration.
We therefore omit this configuration to ensure consistency between the assumed signal evolution timescales and the data availability. The analysis is consequently performed over detector data spanning from 14 January 2025 to 15 March 2025 (GPS: 1420910601--1426068321) for GW250114 and from 7 February 2025 to 27 March 2025 (GPS: 1422988503--1427071863) for GW250207.

\begin{figure*}[t]
 \centering
 \subfigure[GW250114]{\includegraphics[width=0.45\linewidth]{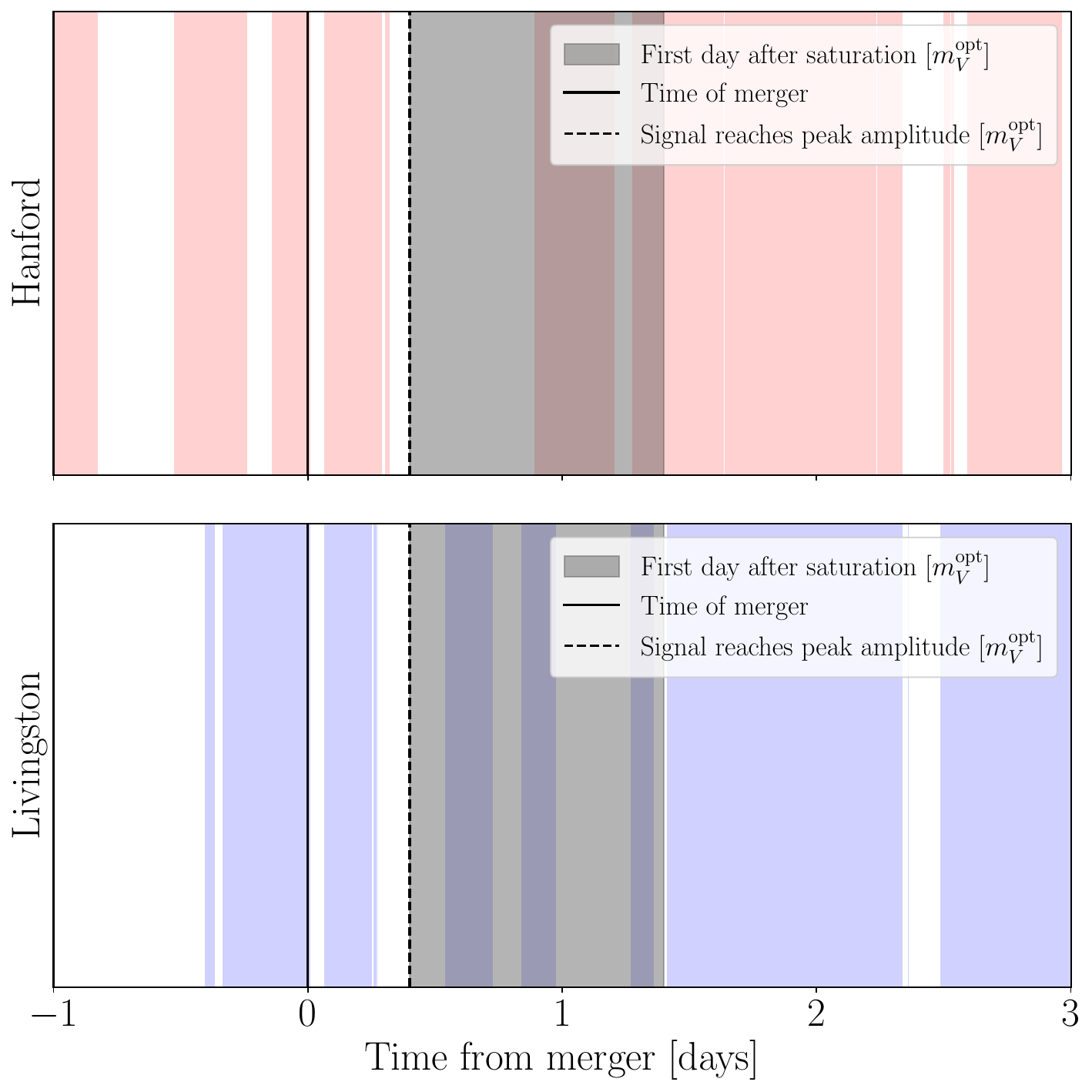} \label{fig:S250114_datagap}}\qquad
  \subfigure[GW250207]{\includegraphics[width=0.45\linewidth]{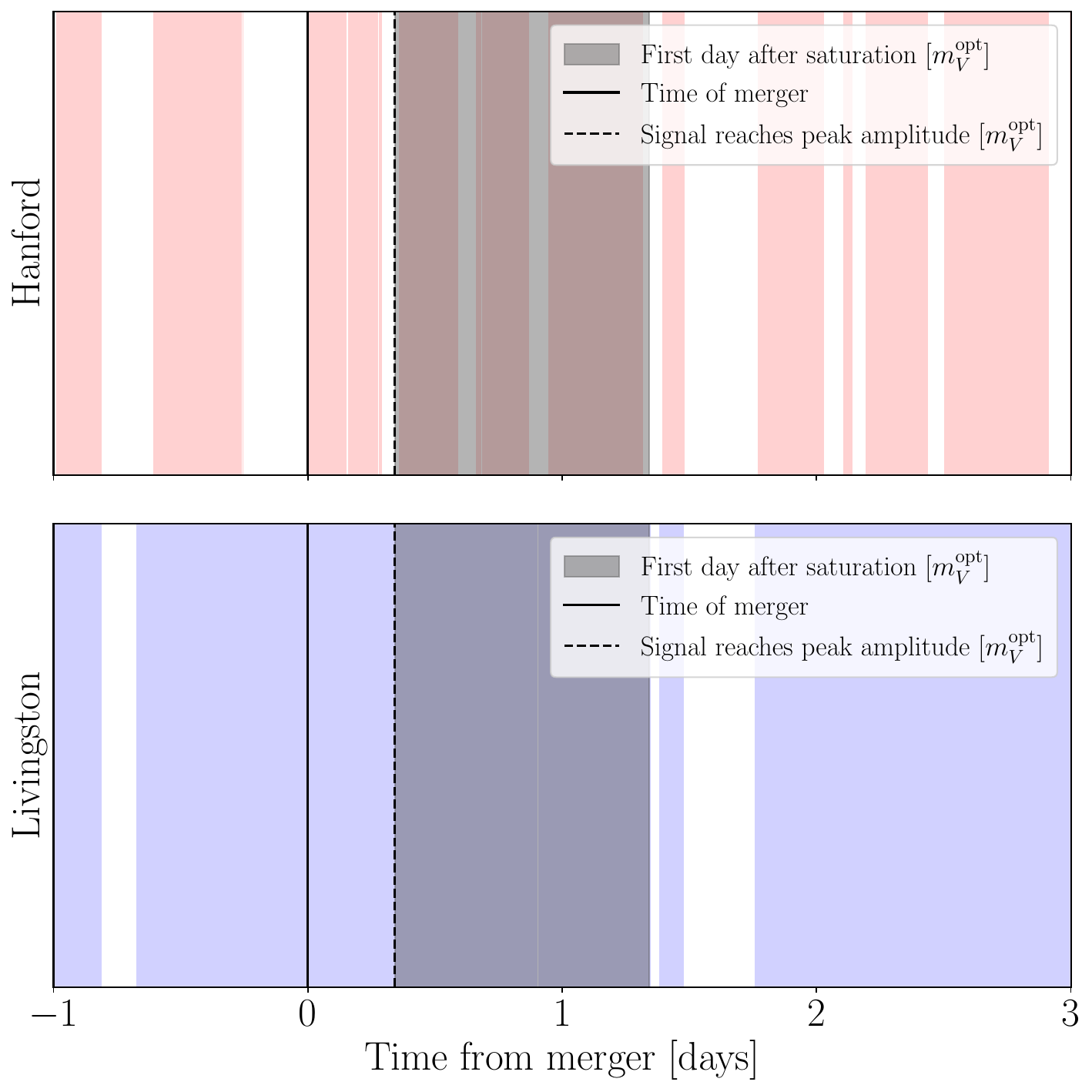}\label{fig:S250207_datagap}} 
  \caption{Data gaps in the LIGO Hanford and Livingston detectors relative to the merger times of GW250114 (left) and GW250207 (right). %\WE{The red/blue regions correspond to where there is data as opposed to where there is a gap, right? Might be slightly ambiguous since the caption reads gap.}
  The red and blue shaded regions show the available data from Hanford and Livingston, respectively, while unshaded intervals indicate data gaps.
  The solid vertical line marks the merger time, while the dotted lines indicate the times at which the boson-cloud signal reaches peak amplitude for the optimal boson mass. The gray shaded region denotes the following one-day interval for the same optimal-boson-mass case. Time is shown in days relative to the merger.}
  \label{fig:data_gap}
\end{figure*}

Over the full selected analysis intervals, the duty cycle for both events is $\sim 60$--$70\%$ in each detector.
The HMM search itself naturally accommodates data gaps. During intervals with no available observational data, equal likelihoods are assigned to all frequency bins~\cite{Jones2023}.
However, the temporal distribution of data availability differs between the two events, particularly in the early post-merger phase. 
In the first day after the expected superradiance saturation time, when the signal reaches its maximum amplitude for $m_V^{\rm opt}$, the duty cycle differs significantly between the events. 
For GW250114, the duty cycle is 44\% in Hanford and 42\% in Livingston, indicating intermittent coverage in either detector during the epoch of expected maximum signal strength and reducing the sensitivity of configurations targeting the shortest signals. 
In contrast, GW250207 has substantially higher coverage, with duty cycles of 80.5\% in Hanford and 99.7\% in Livingston in the first day, corresponding to nearly continuous Livingston data and only limited gaps in Hanford, thereby reducing sensitivity loss during the same epoch. 
Data gaps around the merger time and over the subsequent three days, together with the peak signal times for $m_V^{\rm opt}$, are shown in Fig.~\ref{fig:data_gap}.

\begin{table}[t]
\centering
\setlength{\tabcolsep}{2pt} % default ~6pt
\caption{Search configurations used in the HMM analysis of GW250114 (top) and GW250207 (bottom). For each configuration, we list the SFT duration $T_{\mathrm{SFT}}$, coherence time $T_{\mathrm{coh}}$, total observation time $T_{\mathrm{obs}}$, GPS start time, and the corresponding percentile.}
\begin{ruledtabular}
\begin{tabular}{ccccc}
\% & $T_{\mathrm{SFT}}$ (min) & $T_{\mathrm{coh}}$ (min) & $T_{\mathrm{obs}}$ (min) & $t_{\rm start}$ (s)\\
\hline
2   & 3.45  & 13.8  & 1214.4   & 1420903701 \\
10  & 3.90  & 15.6  & 1466.4   & 1420907361 \\
20  & 4.30  & 17.2  & 1823.2   & 1420910601 \\
30  & 4.90  & 19.6  & 2548.0   & 1420915521 \\
40  & 6.15  & 24.6  & 3960.6   & 1420926381 \\
50  & 8.25  & 33.0  & 6468.0   & 1420943481 \\
60  & 11.55 & 46.2  & 11134.2  & 1420970601 \\
70  & 16.85 & 67.4  & 20152.6  & 1421013921 \\
80  & 25.75 & 103.0 & 38934.0  & 1421086281 \\
90  & 27.80 & 166.8 & 80898.0  & 1421214441 \\
%98 & 29.20 & 262.8 & 162410.4 & 1421407281 \\
\hline
2   & 2.85  & 11.4  & 946.2    & 1422985983 \\
10   & 3.20  & 12.8  & 1126.4   & 1422988503 \\
20   & 3.50  & 14.0  & 1428.0   & 1422990843 \\
30   & 4.00  & 16.0  & 2000.0   & 1422995463 \\
40   & 5.00  & 20.0  & 3060.0   & 1423004103 \\
50   & 6.70  & 26.8  & 4931.2   & 1423017963 \\
60   & 9.35  & 37.4  & 8452.4   & 1423040043 \\
70   & 14.00 & 56.0  & 15792.0  & 1423078143 \\
80   & 21.15 & 84.6  & 29948.4  & 1423136463 \\
90   & 27.60 & 138.0 & 63756.0  & 1423246503 \\
%98   & 27.25 & 218.0 & 126876.0 & 1423405323 \\
\end{tabular}
\end{ruledtabular}
\label{tab:config}
\end{table}

%%%%%%%%%%%%%%%%%%%%%%%%%%%%%%%%%%%%%%%%%%%%%%%%%%%%%%%%%%%%%%%%%%
\subsection{Candidate follow-up\label{Subsec:follow_up}} 
To distinguish potential signals from non-astrophysical artifacts, we implement a series of follow-up veto techniques following the framework described in Ref.~\cite{LIGOScientific:2025csr}. A brief summary is provided here. All follow-up stages prior to manual inspection, including the initial candidate selection based on a detection-statistic threshold and the known-line and single-interferometer vetoes, are predefined and applied uniformly to all candidates.

In each $1~\mathrm{Hz}$ band and for each search configuration, candidates are required to satisfy a threshold on the detection statistic, $\bar{\mathcal{L}} > \bar{\mathcal{L}}_{\mathrm{th}}$, where the threshold corresponds to a per-band, per-configuration false-alarm probability of 1\%. The value of $\bar{\mathcal{L}}_{\mathrm{th}}$ is estimated empirically for each choice of $T_{\mathrm{coh}}$ using 300 realizations of Gaussian noise, taking the 99th percentile of the resulting distribution. Surviving candidates are then subjected to a known-line veto, in which the recovered Viterbi path is compared against the catalog of known instrumental lines in each detector and candidates with overlapping paths are rejected~\cite{known_lines_O4abc,Goetz2026,LIGOScientific:2025csr}. The remaining candidates are then subjected to a single-interferometer veto, followed by manual inspection.

For GW250114 and GW250207, the search spans 84 and 90 individual $1~\mathrm{Hz}$ bands, respectively, and is carried out over 10 configurations, resulting in 840 and 900 optimal Viterbi paths, respectively. These paths correspond to the maximum-likelihood frequency-evolution tracks returned by the HMM tracking for each configuration in each $1~\mathrm{Hz}$ sub-band~\cite{Jones2023,Jones2025}. 
%A fixed sky location is used for both events. 
After applying the initial threshold cut, the number of candidates is reduced to 52 for GW250114 and 64 for GW250207. 
The known-line veto~\cite{known_lines_O4abc,Goetz2026} further reduces these to 3 and 12 candidates, respectively. 
%Applying the single-interferometer veto eliminates all remaining candidates for GW250114, while 11 candidates persist for GW250207. 
%We then apply a conservative single-interferometer veto, which compares the detection statistics and Viterbi paths obtained from single-detector and multi-detector searches to identify candidates whose significance is enhanced in one detector, with no corresponding contribution from the other detector. 
We then apply a conservative single-interferometer veto by repeating the search using each detector individually. A candidate is vetoed if its detection statistic is below threshold in one detector but exceeds that of the multi-detector search in the other detector, and the corresponding Viterbi paths overlap~\cite{LIGOScientific:2025csr}.
Such candidates are therefore consistent with detector-specific instrumental artifacts. This veto eliminates all remaining candidates for GW250114, while 10 candidates persist for GW250207.
These remaining candidates are subsequently examined through manual inspection of the spectrograms from each detector in the relevant frequency band and time segments, to determine whether they overlap with any visible noise artifacts. All 10 candidates are found to coincide with detector-specific noise artifacts, predominantly in Hanford, and are therefore vetoed; see Appendix~\ref{app:manual_veto} for details of the manual inspection procedure and the spectrogram generation.
A summary of the candidate counts at each stage of the follow-up procedure is provided in Table ~\ref{tab:candidates_followup}, and a complete list of the vetoed candidates is provided in Ref.~\cite{O4bcDCC}.

\begin{table}[t]
\centering
\caption{Summary of candidate counts after each stage of the follow-up procedure for GW250114 and GW250207.}
\label{tab:candidates_followup}
\renewcommand{\arraystretch}{1.5}
\setlength{\tabcolsep}{8pt}
\begin{tabular}{lcc}
\hline\hline
Stage & \multicolumn{2}{c}{{Remaining candidates}} \\
 & {GW250114} & {GW250207} \\
\hline
Initial candidates   & 52 & 64 \\
Known-line veto      & 3  & 12 \\
Single-IFO veto      & 0  & 10 \\
Manual inspection    & 0  & 0 \\
\hline\hline
\end{tabular}
\end{table}

%%%%%%%%%%%%%%%%%%%%%%%%%%%%%%%%%%%%%%%%%%%%%%%%%%%%%%%%%%%%%%%%%%%%%%%%%%%%%%%%
\subsection{Disfavored parameter space}
\label{Subsec:directed_searches_constraints}

In the absence of detection, we derive the disfavored vector boson mass range using the method outlined in Ref.~\cite{LIGOScientific:2025csr}. We estimate the search sensitivity via simulations. Synthetic signals are generated and injected into Gaussian noise realizations constructed to match real data, with amplitude spectral densities derived from detector data over the corresponding $T_{\rm obs}$ and 1 Hz sub-band (calculated from the harmonic mean over all the $T_{\rm SFT}$ segments), and with data gaps reproduced to reflect those present during the analysis period. 
Because these simulations use matched Gaussian noise realizations rather than injections into the exact measured detector noise, we refer to the resulting mass range as ``disfavored'' rather than strictly ``excluded.''
We also avoid quoting conventional continuous-wave upper limits, since the search does not have a single, well-defined minimum detectable signal amplitude; the recovery probability depends on the remnant black-hole posteriors, the assumed boson mass, and the search configurations. 
%The disfavored boson mass range is most robust in frequency bands where the data are well approximated by Gaussian noise.
For each assumed vector boson mass, we compute the detection probability by averaging the recovery efficiency over both noise realizations and posterior samples of the remnant black hole parameters, thereby accounting for their uncertainties.
Specifically, following Ref.~\cite{LIGOScientific:2025csr}, we draw $N_{\rm samp}=200$ samples from the posterior distribution of the remnant black hole parameters. For each posterior sample and assumed vector boson mass, a synthetic signal is generated and injected into $N_{\rm noise}=10$ independent noise realizations. The recovery efficiency is then averaged over all posterior samples and noise realizations to obtain the detection probability~\cite{Jones2025},
\begin{equation}
P_{\rm det}(m_V)=\frac{1}{N_{\rm samp}N_{\rm noise}}
\sum_{i=1}^{N_{\rm samp}}
N_{\rm det}(\theta_i;m_V),
\end{equation}
where $\theta_i$ denotes a sampled set of black hole parameters and $N_{\rm det}(\theta_i;m_V)$ is the number of recovered signals out of the $N_{\rm noise}$ noise realizations for that sample. Simulations are performed for all search configurations used in the analysis, and a signal is considered recovered if at least one configuration yields a detection statistic above the chosen threshold.
Only this threshold-based candidate selection is applied to the simulations; the subsequent follow-up vetoes used in the search are not included. 
Some persistent narrow-band instrumental lines fall within the frequency ranges of both searches and appear as localized non-Gaussian features, although they correspond to slightly different boson masses because of the different source redshifts.
The resulting detection probabilities may therefore be somewhat optimistic in frequency bands affected by known lines and non-Gaussian noise artifacts, and the quoted disfavored mass ranges should be interpreted as most robust where the detector noise is well approximated by Gaussian noise.

\begin{figure}[] 
\centering
\includegraphics[width=0.5\textwidth]{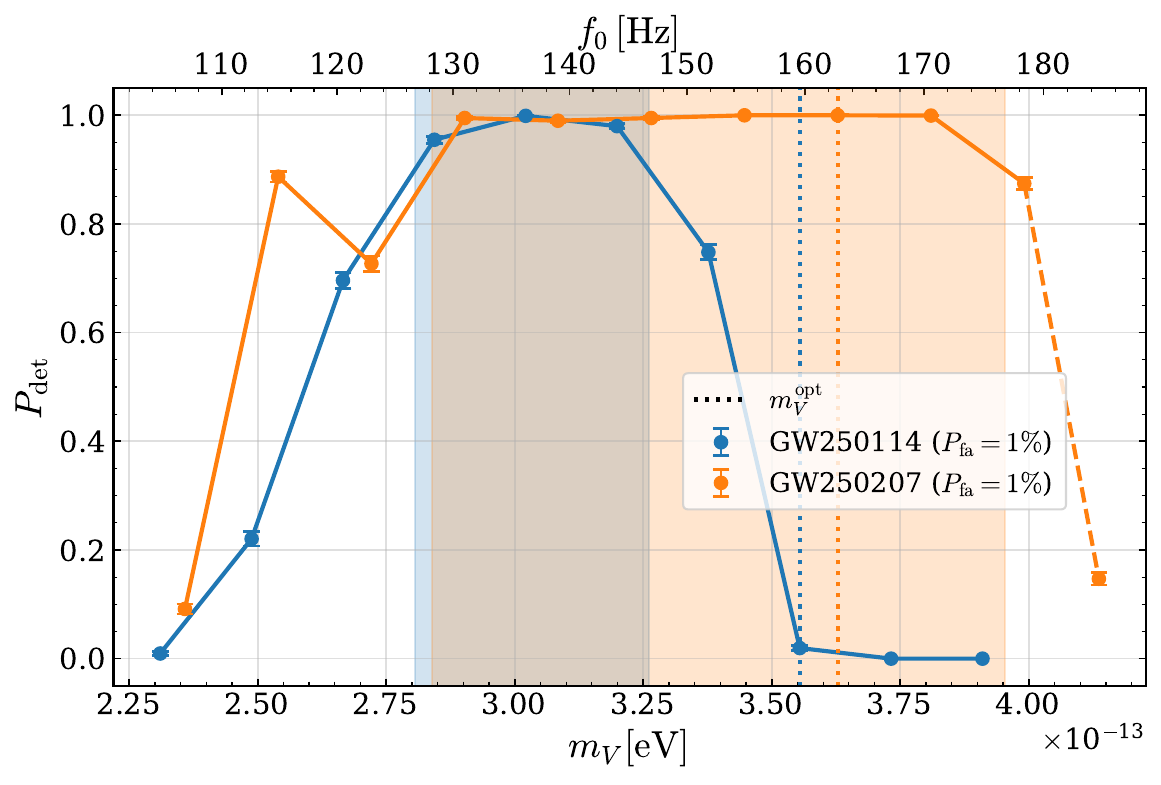}
\caption{Detection probability $P_{\rm det}$ as a function of vector boson mass $m_V$ (bottom axis) and the corresponding gravitational wave frequency in the detector frame (top axis), with a $1\%$ false alarm probability ($P_{\rm fa}$), for the remnant black holes formed in GW250114 (blue) and GW250207 (orange). The vertical dotted lines mark the optimal boson masses $m_V^{\rm opt}$ for the remnants with median parameters listed in Table~\ref{tab:hmm_targets}. The shaded region corresponds to the disfavored boson mass range with greater than 90\% confidence. The dashed segment indicates a simulation-only extension beyond the searched mass range for GW250207, included to verify that $P_{\rm det}$ decreases at higher boson masses. The error bars represent the $1\sigma$ beta-binomial uncertainty in $P_{\rm det}$~\cite{beta-binomial}; some are smaller than the marker size and are therefore not visible.
(The top-axis frequencies are obtained by averaging $f_0$ over search configurations for each boson mass, using the median black-hole parameters in Table~\ref{tab:hmm_targets}; the two events give nearly identical frequency mappings.)}
\label{fig:constraints}
\end{figure}

The resulting detection probability $P_{\rm det}(m_V)$ quantifies the likelihood that a signal would have been detected by our pipeline if present. 
In the absence of detection, this quantity can be interpreted as the confidence level at which a given boson mass is disfavored. 
Although the two events probe similar vector-boson mass ranges, GW250207 yields strong confidence near $m_V^{\rm opt}$, where the expected signals are the shortest, whereas GW250114 lacks sensitivity to these shortest-duration signals due to data gaps during the corresponding early-time interval.
For a false alarm probability of $P_{\rm fa} = 1\%$, we disfavor vector boson mass ranges of \vectorlimitssearchone eV for GW250114 and \vectorlimitssearchtwo eV for GW250207 at 90\% confidence, as shown in Fig.~\ref{fig:constraints}. We quote a combined interval of \vectorlimitssearch~eV, defined as the union of these independently disfavored mass ranges rather than as the result of a joint statistical combination of the two searches.
We also include one simulation-only point above the upper boundary of the searched mass range for GW250207 (connected by the dashed line) to verify the expected decrease in $P_{\rm det}$ at higher vector-boson masses. This point is not part of the search and is not used to define the disfavored mass range.

These results can be compared with those obtained in the O4a analysis of Ref.~\cite{LIGOScientific:2025csr} (see Fig.~3 therein), where the disfavored mass ranges for GW230814 and GW231123 were reported at a lower confidence level of $\sim30\%$ for $P_{\rm fa} = 1\%$. The higher confidence achieved with GW250114 and GW250207 here is primarily driven by their closer distance, higher signal-to-noise ratios, and tighter constraints on source parameters.
Improved sky localization reduces the parameter space explored by the search, while smaller luminosity distances increase the expected signal amplitude. In particular, GW250207 is located at a distance of $\sim 187$~Mpc, significantly closer than typical O4a events~\cite{GWTC-4_results}, leading to a higher recovery efficiency and consequently stronger constraints on the vector boson mass.

%%%%%%%%%%%%%%%%%%%%%%%%%%%%%%%%%%%%%%%%%%%%%%%%%%%%%%%%%%%%%%%%%%%%%%%%%%%%%%%%
\section{Constraints from pre-merger black holes}
\label{Sec:spin_measurements_constraints}

\subsection{Method}
\label{Subsec:spin_constrain_method}

In addition to the directed HMM searches, we implement a complementary method to constrain ultralight bosons (scalar and vector) using spin measurements of binary black hole merger events, following Ref.~\cite{aswathi2025}. This method exploits the fact that superradiant instabilities impose an upper bound on black hole spin for a given boson mass.
If ultralight bosons of mass $m_b$ exist, black holes can undergo superradiant spin-down over a characteristic timescale $\tau_{\rm growth}(M,\chi,m_b)$. For systems with age $T_{\rm age} \gtrsim \tau_{\rm growth}$, this leads to a maximum allowed spin $\chi_{\max}(M, m_b, T_{\rm age})$, defining an excluded region in the mass–spin plane.

We compute $\chi_{\text{max}}$ using the \texttt{SuperRad} package ~\cite{May2024,Siemonsen2023}, assuming conservatively that the cloud grows from vacuum fluctuations to the saturation mass at which the instability shuts off.  We include fully relativistic superradiant frequencies and instability rates for dominant modes with $m \leq 2$ and adopt nonrelativistic approximations for higher modes as described in Ref.~\cite{aswathi2025}. 

The efficiency of superradiant spin-down depends on the elapsed time between black-hole formation and merger. Following Refs.~\cite{aswathi2025,Abac_2025_GW241011_241110}, we consider $T_{\text{age}} \in [10^5,10^7]$ yr, where the lower bound is motivated by merger timescales for dynamically formed binaries in a dense environment, and the upper bound is adopted as a representative inspiral timescale for isolated binary evolution~\cite{PortegiesZwart:1999nm,Dominik:2013tma,Bavera:2020inc}. 
The exclusion region expands with $T_{\text{age}}$, since older black holes would have undergone superradiant spin-down for a longer time, thereby strengthening the constraints on the boson mass, although the dependence is not strong over this range.
%Unless otherwise stated, we adopt $T_{\text{age}} = 10^5$ yr when quoting conservative constraints. 
We neglect possible perturbative effects of the binary companion on cloud growth, so $T_{\text{age}}$ should be interpreted as a lower bound on the time available for superradiance prior to binary interaction suppressing the instability. 

For each merger event, we use posterior samples from the event parameter estimation. Denoting the primary black-hole mass and dimensionless spin by $m_1^{(i)}$ and $\chi_1^{(i)}$ we compute the posterior-driven nonexclusion probability
\begin{equation}
   P(m_b, T_{\rm age}) = \frac{1}{N_{\mathrm{samp}}} \sum_{i=1}^{N_{\mathrm{samp}}} I{\left\{ \chi_1^{(i)} < \chi_{\max}(m_1^{(i)}, m_b, T_{\rm age}) \right\}},
   \label{eq:nonexclusion}
\end{equation}
where $I\left\{\cdot \right\}$ is an indicator function that returns unity if the primary BH spin lies below its respective superradiance-imposed upper bound $\chi_{\max}$, and zero otherwise, and $N_{\rm samp} \sim \mathcal{O}(10^4)$ is the number of posterior samples. 
%as secondary spins are less well constrained for these events and do not contribute meaningfully to the resulting constraints. 
We restrict the reported constraints to the primary components, though in principle, information about the secondary spin could extend the constraints. For the systems considered here, the secondary spin posteriors remain close to their priors and are largely uninformative. Applying the exclusion criterion defined below to the secondary components individually yields no excluded boson-mass intervals for the black-hole ages considered here. Jointly using both component spins while taking into account their correlations could still strengthen the constraints, though we conservatively do not consider that here.

We also account for the possibility that the exclusion could be influenced by the spin prior rather than by the data. 
In some regions of parameter space, the maximum allowed spin after superradiant spin-down, $\chi_{\max}$, can be sufficiently small that it removes a non-negligible fraction of the prior support. 
To quantify this effect, we compute a prior-driven non-exclusion probability, $P'(m_b,T_{\rm age})$, using the same procedure as for $P(m_b,T_{\rm age})$, but drawing $\chi_1$ directly from the uniform spin prior (consistent with the spin priors adopted in GWTC-5~\cite{GWTC-5-method}). 
We then regard a boson mass $m_b$ as excluded at least at 90\% confidence for a given spin-down timescale $T_{\rm age}$ only if
\begin{equation}
    P(m_b,T_{\rm age}) < 0.1\,P'(m_b,T_{\rm age}) .
\end{equation}
This requirement is more conservative than the nominal criterion $P(m_b,T_{\rm age})<0.1$, because it demands that the posterior-based non-exclusion probability be reduced by more than a factor of 10 relative to the prior-only expectation. 
In this way, the inferred constraint is required to be driven by the measured spin information rather than by the assumed uniform spin prior. More details are provided in Ref.~\cite{aswathi2025}.

%%%%%%%%%%%%%%%%%%%%%%%%%%%%%%%%%%%%%%%%%%%%%%%%%%%%%%%%%%%%%%%%%%%%%%%%%%%%%%%%
\subsection{Results}
\label{Subsec:spin_constraint_results}
\begin{figure}[] 
\centering
\includegraphics[width=0.5\textwidth]{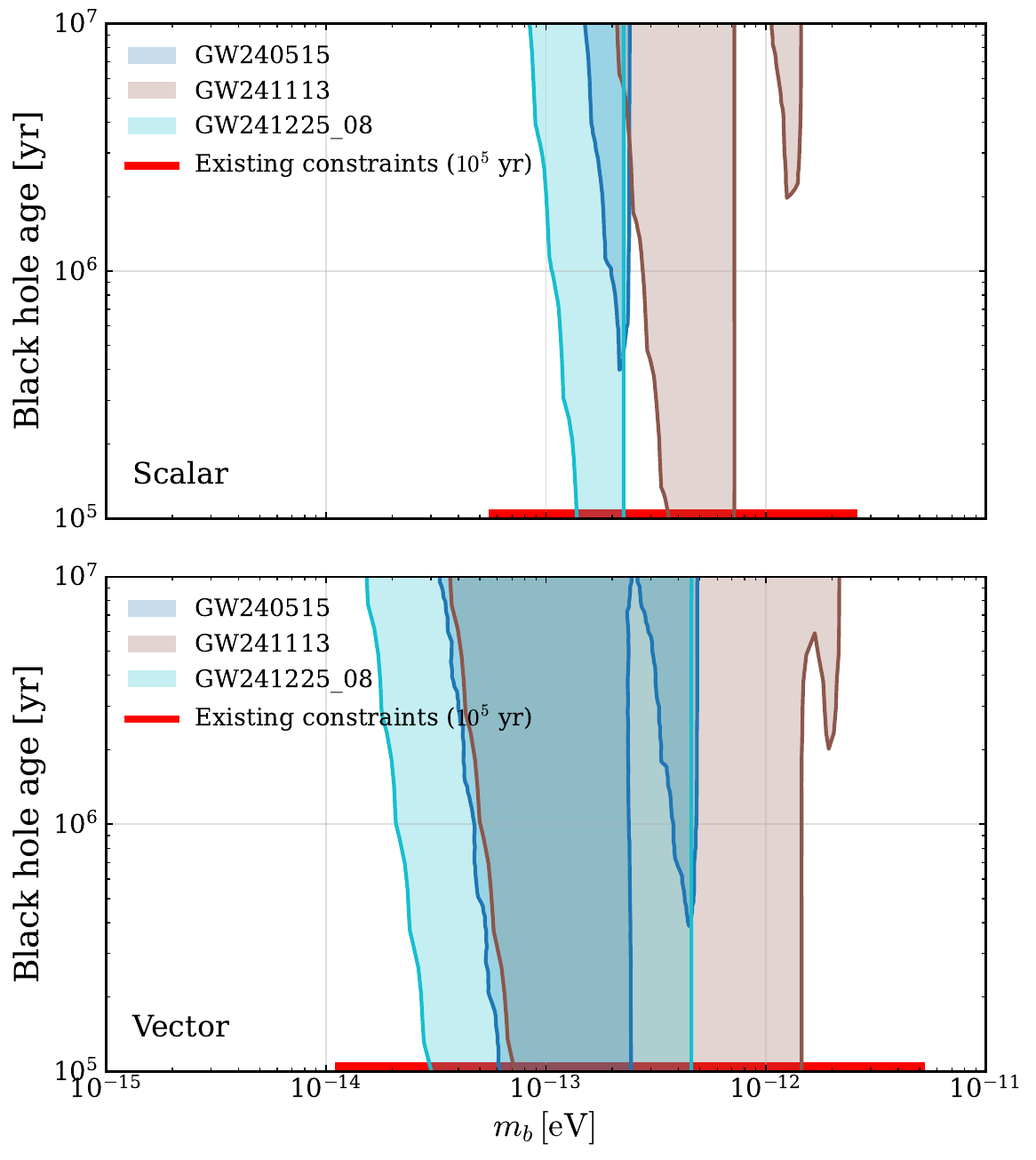}
\caption{Exclusion regions for scalar and vector boson masses
as a function of the black hole age, inferred from the primary black hole spin measurements in the three binaries. The shaded regions are excluded by each event at confidence levels above 90\% (i.e., $P < 0.1P'$)~\cite{aswathi2025}. The red line indicates the conservative boson mass range previously excluded using the same technique~\cite{Abac_2025_GW241011_241110,aswathi2025}, assuming a black hole lifetime of $T_{\rm age} = 10^5$ yr.}
\label{spin_constrain_result}
\end{figure}

We apply the spin-based superradiance framework described in Sec. \ref{Sec:spin_measurements_constraints} to three O4b events: GW240515, GW241113, and GW241225\_08. The resulting exclusion regions in boson mass are shown in Fig. \ref{spin_constrain_result}, where shaded areas correspond to parameters excluded at 90\% confidence. Separate analyses are performed for scalar and vector bosons. Assuming a black-hole age of $T_{\rm age} = 10^5$ yr, we exclude scalar bosons in the mass range \scalarlimitsspin~eV and vector bosons in the mass range \vectorlimitsspin~eV at 90\% confidence. These quoted intervals are constructed as the union of the event-by-event excluded mass ranges from the three O4b events shown in Fig.~\ref{spin_constrain_result}. The excluded region broadens with increasing assumed black-hole age.
 
The excluded mass ranges obtained here are broadly consistent with those previously reported for individual high-spin events in earlier studies~\cite{aswathi2025,Abac_2025_GW241011_241110} (marked by red lines in Fig. \ref{spin_constrain_result}, assuming a black-hole age of $T_{\rm age} = 10^5$ yr).
The stronger constraints from previous events arise from a combination of better-constrained spins for lower-mass black holes and the inclusion of higher-mass black holes, which extend the range of boson masses that can be excluded.
This is expected because these constraints are driven primarily by the constituent black-hole masses and spins, rather than by detector sensitivity alone. The events with well-constrained high spins considered so far have mostly fallen within a similar black-hole mass range and therefore probe a similar boson mass range where the detectors are most sensitive.
As additional high-spin mergers are observed with a wider spread in black hole masses, the excluded region in boson parameter space will continue to expand.
%The vector-boson exclusions obtained from these three events improve the coverage toward higher boson masses compared with the disfavored mass ranges from the directed searches presented in Sec.~\ref{Sec: HMM_search} of this work, while maintaining high confidence at $\sim 10^{-13}$ eV. 
The vector-boson exclusions obtained from these three events span a broader boson mass range than the disfavored region from the directed searches presented in Sec.~\ref{Sec: HMM_search}, including that region while extending to lower and higher boson masses. Both approaches strongly disfavor vector bosons with masses $\sim 10^{-13}$~eV.

%%%%%%%%%%%%%%%%%%%%%%%%%%%%%%%%%%%%%%%%%%%%%%%%%%%%%%%%%%%%%%%%%%%%%%%%%%%%%%%%
\section{Conclusions}
\label{Sec:conclusions}

In this study, we present two complementary approaches to probe ultralight bosons using gravitational-wave observations: directed searches for vector boson cloud emission from merger remnants and constraints derived from high-spin measurements of pre-merger constituent black holes (for scalar and vector bosons). These methods yield consistent and complementary constraints on the boson parameter space.

The directed HMM searches target newly formed merger remnants, where boson clouds may form and emit gravitational waves on timescales of hours to months. No evidence of the signal is found in searches targeting the remnants of two loud merger events, GW250114 and GW250207. The disfavored vector boson mass ranges are derived with high confidence around $3\times 10^{-13}$ eV. These results demonstrate the increasing sensitivity of directed searches as detector performance improves and louder events become available.

The spin-based analysis provides an independent constraint by leveraging the absence of superradiant spin-down in highly spinning pre-merger black holes. 
By analyzing three O4b events, we obtain exclusion regions for both scalar and vector bosons that are consistent with previous constraints from high-spin constituent black holes. The vector-boson exclusions are also broadly consistent with the disfavored range from the directed searches in this study, while extending to higher vector-boson masses.

The two approaches are complementary in both methodology and astrophysical interpretation. The directed search probes the presence of boson clouds around recently formed black holes, with minimal dependence on assumptions about binary evolution, whereas the spin-based constraints rely on the long-term evolution of black holes and are subject to astrophysical uncertainties. 
Therefore, the two methods, while covering overlapping regions of parameter space, provide complementary observational probes of vector bosons based on distinct assumptions.

Continued improvements in detector sensitivity, together with the accumulation of high signal-to-noise ratio events, will enhance the reach of both approaches. In particular, the identification of additional nearby merger remnants and highly spinning black holes will further expand the excluded regions in boson parameter space. Combined analyses that integrate multiple events and jointly model boson cloud dynamics and spin evolution may offer a powerful avenue for future studies.
 
%%%%%%%%%%%%%%%%%%%%%%%%%%%%%%%%%%%%%%%%%%%%%%%%%%%%%%%%%%%%%%%%%%%%%%%%%%%%%%%%

\acknowledgments{}
% updated March 2024
%
% If not using the driver to create stand-alone pdfs for the LVC and LVK, uncomment the following lines to create the sample pdf with options spelled out
\newif\ifcoreonly\coreonlytrue
\newif\ifkagra\kagratrue
\newif\ifheader\headerfalse
\ifheader
\begin{center}{\bf\Large
\ifkagra
Conference proceedings acknowledgements for \\ the LIGO Scientific Collaboration, the Virgo Collaboration and the KAGRA Collaboration
\else
Conference proceedings acknowledgements for \\ the LIGO Scientific Collaboration and the Virgo Collaboration
\fi
}\end{center}
\fi
This material is based upon work supported by NSF's LIGO Laboratory, which is a
major facility fully funded by the National Science Foundation.
The authors also gratefully acknowledge the support of
the Science and Technology Facilities Council (STFC) of the
United Kingdom, the Max-Planck-Society (MPS), and the State of
Niedersachsen/Germany for support of the construction of Advanced LIGO 
and construction and operation of the GEO\,600 detector. 
Additional support for Advanced LIGO was provided by the Australian Research Council.
The authors gratefully acknowledge the Italian Istituto Nazionale di Fisica Nucleare (INFN),  
the French Centre National de la Recherche Scientifique (CNRS) and
the Netherlands Organization for Scientific Research (NWO)
for the construction and operation of the Virgo detector
and the creation and support  of the EGO consortium. 
The authors also gratefully acknowledge research support from these agencies as well as by 
the Council of Scientific and Industrial Research of India, 
the Department of Science and Technology, India,
the Science \& Engineering Research Board (SERB), India,
the Ministry of Human Resource Development, India,
the Spanish Agencia Estatal de Investigaci\'on (AEI),
the Spanish Ministerio de Ciencia, Innovaci\'on y Universidades,
the European Union NextGenerationEU/PRTR (PRTR-C17.I1),
the ICSC - CentroNazionale di Ricerca in High Performance Computing, Big Data
and Quantum Computing, funded by the European Union NextGenerationEU,
the Comunitat Auton\`oma de les Illes Balears through the Conselleria d'Educaci\'o i Universitats,
the Conselleria d'Innovaci\'o, Universitats, Ci\`encia i Societat Digital de la Generalitat Valenciana and
the CERCA Programme Generalitat de Catalunya, Spain,
the Polish National Agency for Academic Exchange,
the National Science Centre of Poland and the European Union - European Regional
Development Fund;
the Foundation for Polish Science (FNP),
the Polish Ministry of Science and Higher Education,
the Swiss National Science Foundation (SNSF),
the Russian Science Foundation,
the European Commission,
the European Social Funds (ESF),
the European Regional Development Funds (ERDF),
the Royal Society, 
the Scottish Funding Council, 
the Scottish Universities Physics Alliance, 
the Hungarian Scientific Research Fund (OTKA),
the French Lyon Institute of Origins (LIO),
the Belgian Fonds de la Recherche Scientifique (FRS-FNRS), 
Actions de Recherche Concert\'ees (ARC) and
Fonds Wetenschappelijk Onderzoek - Vlaanderen (FWO), Belgium,
the Paris \^{I}le-de-France Region, 
the National Research, Development and Innovation Office of Hungary (NKFIH), 
the National Research Foundation of Korea,
the Natural Sciences and Engineering Research Council of Canada (NSERC),
the Canadian Foundation for Innovation (CFI),
the Brazilian Ministry of Science, Technology, and Innovations,
the International Center for Theoretical Physics South American Institute for Fundamental Research (ICTP-SAIFR), 
the Research Grants Council of Hong Kong,
the National Natural Science Foundation of China (NSFC),
the Israel Science Foundation (ISF),
the US-Israel Binational Science Fund (BSF),
the Leverhulme Trust, 
the Research Corporation,
the National Science and Technology Council (NSTC), Taiwan,
the United States Department of Energy,
and
the Kavli Foundation.
The authors gratefully acknowledge the support of the NSF, STFC, INFN and CNRS for provision of computational resources.

\ifcoreonly\else
{\bf For papers using O3b (and future) data, the following paragraph should be added for KAGRA.}\\
\fi
\ifkagra
This work was supported by MEXT,
the JSPS Leading-edge Research Infrastructure Program,
JSPS Grant-in-Aid for Specially Promoted Research 26000005,
JSPS Grant-in-Aid for Scientific Research on Innovative Areas 2402: 24103006,
24103005, and 2905: JP17H06358, JP17H06361 and JP17H06364,
JSPS Core-to-Core Program A.\ Advanced Research Networks,
JSPS Grants-in-Aid for Scientific Research (S) 17H06133, 20H05639 and 26K21721,
JSPS Grant-in-Aid for Transformative Research Areas (A) 26A204: JP26H00407,
the joint research program of the Institute for Cosmic Ray Research,
University of Tokyo,
the National Research Foundation (NRF),
the Computing Infrastructure Project of the Global Science experimental Data hub
Center (GSDC) at KISTI,
the Korea Astronomy and Space Science Institute (KASI),
the Ministry of Science and ICT (MSIT) in Korea,
Academia Sinica (AS),
the AS Grid Center (ASGC) and the National Science and Technology Council (NSTC)
in Taiwan under grants including the Science Vanguard Research Program,
the Advanced Technology Center (ATC) of NAOJ, 
the Mechanical Engineering Center of KEK
and Vietnam National Foundation for Science and Technology Development 
(NAFOSTED) 103.01-2025.147.
\fi

Additional acknowledgements for support of individual authors may be found in the following document: \\
\texttt{https://dcc.ligo.org/LIGO-M2300033/public}.
For the purpose of open access, the authors have applied a Creative Commons Attribution (CC BY)
license to any Author Accepted Manuscript version arising.
% Update next sentence to another first author in the future, as needed
% and change "LIGO-Virgo-KAGRA" to "LIGO-Virgo" for any remaining O3a papers
We request that citations to this article use 'A. G. Abac {\it et al.} (LIGO-Virgo-KAGRA Collaboration), ...' or similar phrasing, depending on journal convention.

\ifcoreonly\else
{\bf For collaboration papers accepted after revisions made in response to referees, it is recommended that the 
following sentence be included: (with appropriate choice of singular or plural):}\\
We thank the anonymous journal referee(s) for helpful comments.
\fi

\ifcoreonly\else
{\bf For certain collaboration papers, it may be appropriate to acknowledge specific analysis software.
One template to consider is that used for the GW190425 discovery paper, for which the latex can be
found here:}\\
{\small https://git.ligo.org/publications/gw190425/gw190425-discovery/-/blob/master/gw190425-discovery.tex\#L259.}
\fi

\ifcoreonly\else
{\bf For collaboration papers released after March 2020, it may be appropriate to add the following special acknowledgement, BUT the journal may reject it (PRL rejected it). If one chooses, one can include such acknowledgements in arXiv/DCC versions only in that case:}\\
\fi
\ifcoreonly\else
{\it We would like to thank all of the essential workers who put their health at risk during the COVID-19 pandemic, without whom we would not have been able to complete this work.}
\fi

\appendix
\section{Vector boson signal model for remnant black holes}
\label{app:vector_waveform}

Using the \texttt{SuperRad} waveform model~\cite{May2024,Siemonsen2023} and the median remnant masses and spins listed in Table~\ref{tab:hmm_targets}, we determine, for each system, the vector boson mass that maximizes the superradiant instability. This optimal mass, $m_V^{\rm opt}$, yields the largest gravitational-wave strain amplitude attainable by a given black hole-boson system at a reference time $t_{\rm ref}$, defined as the time at which the boson cloud reaches its full growth and the instability saturates.

For each remnant, \texttt{SuperRad} is used to compute the corresponding gravitational-wave signal parameters, including the cloud growth timescale $\tau_{\rm growth}$, the emission (depletion) timescale $\tau_{\rm GW}$, the strain amplitude $h_0$, the time-dependent emission frequency $f_{\rm GW}$ and the frequency derivative $\dot{f}_{\rm GW}$. 

The initial gravitational frequency in the source frame is set by the angular frequency of the saturated cloud, $f=\omega/\pi$, while its subsequent evolution can be approximated as,
\begin{equation}
    f_{\rm GW}(t) = f_{\infty } - \frac{|f_{\text{shift}}|}{1+(t-t_{\text{sat}})/\tau_{\text{GW}}},
\end{equation}
where $ f_{\infty }$ is the asymptotic frequency at late times and $ f_{\text{shift}}$ accounts for the initial self-gravity of the cloud. In practice, \texttt{SuperRad} incorporates higher-order relativistic corrections to compute $f_{\rm GW}$ and $\dot{f}_{\rm GW}$ with high accuracy. 

As both search targets are merger remnants at cosmological distances, we account for redshift effects when mapping to the detector frame. The observed gravitational-wave frequency and its derivative scale as
\begin{equation}
f_{\rm det} = \frac{f_{\rm src}}{1 + z}, \qquad
\dot{f}_{\rm det} = \frac{\dot{f}_{\rm src}}{(1 + z)^2},
\end{equation}
while the emission timescale is correspondingly dilated,
\begin{equation}
\tau_{\rm GW}^{\rm det} = (1 + z)\,\tau_{\rm GW}^{\rm src},
\end{equation}
where the superscripts `src' and `det' denote the source and detector frames, respectively.

\section{Manual inspection of final candidates}
\label{app:manual_veto}
The manual inspection is performed using spectrograms generated from calibrated strain data. The spectrograms are constructed using the Welch method by averaging 60-s Fourier transforms over the inspected duration, corresponding to a frequency resolution of approximately 0.017~Hz. To ensure consistency with the HMM search, only analysis-ready data segments included in the search are used to construct the spectrograms.

All 10 remaining candidates for GW250207 overlap with noise artifacts visible in the spectrograms, predominantly in the Hanford detector. The surviving candidates' starting frequencies, search configurations, and detection statistics are listed in Table~\ref{tab:manual_candidates}. In this section, we present representative examples illustrating the basis for their veto. 

For each candidate, spectrograms from the Hanford and Livingston detectors are examined in the relevant frequency sub-band and time segment. A common pattern is observed across all surviving candidates: the candidate tracks coincide with visible wandering noise features in the Hanford data, while the corresponding Livingston spectrograms remain comparatively free of similar features.
%the candidate tracks coincide with prominent excess-power features in the Hanford data, while the corresponding Livingston spectrograms remain comparatively free of similar disturbances. 
Furthermore, the optimal paths recovered from the combined Hanford--Livingston data generally lie close to those from the Hanford data alone, indicating that the candidate significance is dominated by detector-specific artifacts rather than coherent excess power present in both detectors.

\begin{figure*}[] 
\centering
\includegraphics[width=\textwidth]{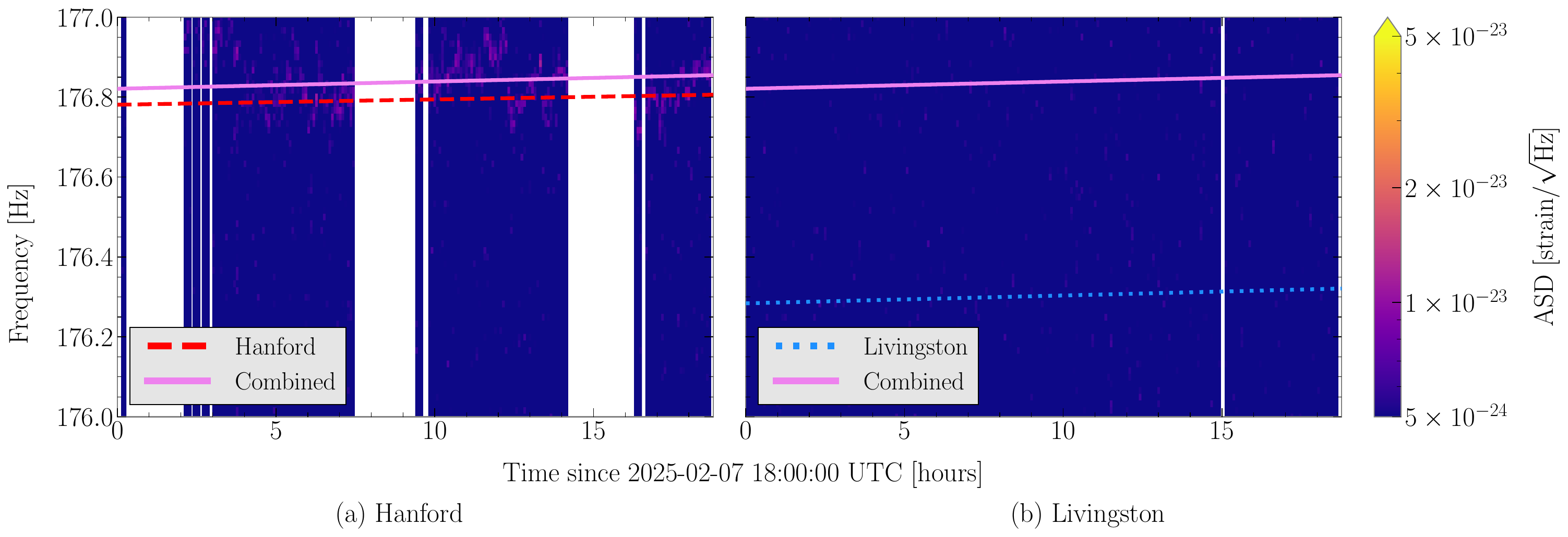}
\caption{Representative example of a candidate vetoed during manual inspection. The left and right panels show the Hanford and Livingston spectrograms, respectively. The red dashed, blue dotted, and magenta solid tracks denote the optimal paths obtained from Hanford-only, Livingston-only, and combined Hanford--Livingston data, respectively. Gaps in the spectrograms correspond to intervals that were not analysis-ready and were therefore excluded from the search.}
\label{fig:spectrogram} 
\end{figure*}

Figure~\ref{fig:spectrogram} shows a representative example corresponding to a candidate with $f_0 = 176.82~\mathrm{Hz}$, $t_{\rm start} = 1422988503$~s, $T_{\mathrm{coh}} = 12.8~\mathrm{min}$, and $T_{\mathrm{obs}}\approx 19~\mathrm{h}$. A weak noise track can be seen near the recovered track, whereas no comparable structure is present in the Livingston data. The optimal path obtained by combining data from both detectors closely follows the Hanford-only path, consistent with a detector-specific noise artifact. Similar behavior is observed for the remaining candidates. The gaps visible in the spectrograms correspond to intervals that were not analysis-ready and were therefore excluded from both the HMM search and the spectrogram.

\begin{table}[t]
\centering
\caption{Start frequencies, search configurations, and detection statistics of the 10 candidates subjected to manual inspection for GW250207. 
}
\label{tab:manual_candidates}

\renewcommand{\arraystretch}{1.5}
\setlength{\tabcolsep}{5pt}

\begin{tabular}{ccccc}
\hline\hline
$f_0$ (Hz) & $t_{\rm start}$ (s) & $T_{\rm coh}$ (min) & $T_{\rm obs}$ (min) & $\bar{\mathcal{L}}$ \\
\hline
176.81 & 1422985983 & 11.4 & 946.2   & 7.48 \\
178.97 & 1422985983 & 11.4 & 946.2   & 7.54 \\
145.04 & 1422988503 & 12.8 & 1126.4  & 8.48 \\
176.82 & 1422988503 & 12.8 & 1126.4  & 7.51 \\
178.97 & 1422988503 & 12.8 & 1126.4  & 7.47 \\
145.04 & 1422990843 & 14.0 & 1428.0  & 8.08 \\
176.79 & 1422990843 & 14.0 & 1428.0  & 7.42 \\
178.63 & 1422990843 & 14.0 & 1428.0  & 7.23 \\
%145.18 & 1422995463 & 16.0 & 2000.0  & 6.93 \\
176.79 & 1423004103 & 20.0 & 3060.0  & 7.24 \\
144.97 & 1423136463 & 84.6 & 29948.4 & 6.71 \\
\hline\hline
\end{tabular}
\end{table}

%\begin{figure*}[]
%\subfigure[Hanford]{\includegraphics[width=0.45\linewidth]{figures/spectrogram_H1.pdf} \label{fig:spectrogram_H1}}\qquad \subfigure[Livingston]{\includegraphics[width=0.45\linewidth]{figures/spectrogram_L1.pdf}\label{fig:spectrogram_L1}} \caption{Representative example of a candidate vetoed during manual inspection. The left and right panels show the Hanford and Livingston spectrograms, respectively. The red, blue, and magenta tracks denote the optimal paths obtained from Hanford-only, Livingston-only, and combined Hanford--Livingston data, respectively. \todo{placeholder; make them nicer}} \label{fig:spectrogram} 
%\end{figure*}

\bibliography{references.bib}{}

\end{document}